\documentclass{sig-alternate}

\usepackage{times}
\usepackage{amsmath,amssymb}
\usepackage{color}
\usepackage{latexsym}
\usepackage{graphicx}
\usepackage{comment}
\usepackage{hhline}
\usepackage{dsfont}
\usepackage[pdfstartview=FitH]{hyperref} %generate bookmark and reference hyperlink, fit pdf content to the window
\usepackage[open,openlevel=2,numbered]{bookmark}

\begin{document}
%
% --- Author Metadata here ---
%\conferenceinfo{WOODSTOCK}{'97 El Paso, Texas USA}
%\CopyrightYear{2007} % Allows default copyright year (20XX) to be over-ridden - IF NEED BE.
%\crdata{0-12345-67-8/90/01}  % Allows default copyright data (0-89791-88-6/97/05) to be over-ridden - IF NEED BE.
% --- End of Author Metadata ---

\title{DeepGraph: Graph Structure Predicts Network Growth}
%
% You need the command \numberofauthors to handle the 'placement
% and alignment' of the authors beneath the title.
%
% For aesthetic reasons, we recommend 'three authors at a time'
% i.e. three 'name/affiliation blocks' be placed beneath the title.
%
% NOTE: You are NOT restricted in how many 'rows' of
% "name/affiliations" may appear. We just ask that you restrict
% the number of 'columns' to three.
%
% Because of the available 'opening page real-estate'
% we ask you to refrain from putting more than six authors
% (two rows with three columns) beneath the article title.
% More than six makes the first-page appear very cluttered indeed.
%
% Use the \alignauthor commands to handle the names
% and affiliations for an 'aesthetic maximum' of six authors.
% Add names, affiliations, addresses for
% the seventh etc. author(s) as the argument for the
% \additionalauthors command.
% These 'additional authors' will be output/set for you
% without further effort on your part as the last section in
% the body of your article BEFORE References or any Appendices.

\numberofauthors{1} %  in this sample file, there are a *total*
% of EIGHT authors. SIX appear on the 'first-page' (for formatting
% reasons) and the remaining two appear in the \additionalauthors section.
%
\author{
% You can go ahead and credit any number of authors here,
% e.g. one 'row of three' or two rows (consisting of one row of three
% and a second row of one, two or three).
%
% The command \alignauthor (no curly braces needed) should
% precede each author name, affiliation/snail-mail address and
% e-mail address. Additionally, tag each line of
% affiliation/address with \affaddr, and tag the
% e-mail address with \email.
%
% 1st. author
\alignauthor
Cheng Li$^{1}$, Xiaoxiao Guo\footnotemark[1]$^{2}$, Qiaozhu Mei$^{1,2}$\\
		\affaddr{$^{1}$School of Information, University of Michigan, Ann Arbor, MI, USA}\\
		\affaddr{$^{2}$Department of EECS, University of Michigan, Ann Arbor, MI, USA}\\
       \affaddr{\{lichengz, \ guoxiao, \ qmei\}@umich.edu}
}
% There's nothing stopping you putting the seventh, eighth, etc.
% author on the opening page (as the 'third row') but we ask,
% for aesthetic reasons that you place these 'additional authors'
% in the \additional authors block, viz.
% Just remember to make sure that the TOTAL number of authors
% is the number that will appear on the first page PLUS the
% number that will appear in the \additionalauthors section.

\maketitle

\begin{abstract}
The topological (or graph) structures of real-world networks are known to be predictive of multiple dynamic properties of the networks. Conventionally, a graph structure is represented using an adjacency matrix or a set of hand-crafted structural features. These representations either fail to highlight local and global properties of the graph or suffer from a severe loss of structural information. There lacks an effective graph representation, which hinges the realization of the predictive power of network structures. 

In this study, we propose to learn the represention of a graph, or the topological structure of a network, through a deep learning model. This end-to-end prediction model, named DeepGraph, takes the input of the raw adjacency matrix of a real-world network and outputs a prediction of the growth of the network. The adjacency matrix is first represented using a graph descriptor based on the heat kernel signature, which is then passed through a multi-column, multi-resolution convolutional neural network. Extensive experiments on five large collections of real-world networks demonstrate that the proposed prediction model significantly improves the effectiveness of existing methods, including linear or nonlinear regressors that use hand-crafted features, graph kernels, and competing deep learning methods.

\end{abstract}

%% A category with the (minimum) three required fields
%%A category including the fourth, optional field follows...
%\category{D.2.8}{Software Engineering}{Metrics}[complexity measures, performance measures]
%

\section{Introduction}
Today we are surrounded by real-world networks of people, information, and technology. These heterogeneous, large scale, and fast evolving networks have provided a new perspective of scientific research, which has resulted in a rapid development of new theories, algorithms, and applications. 

How to model and predict the dynamic properties of social or information networks has received considerable attentions recently~\cite{szabo2010predicting,yang2010modeling,backstrom2006group,romero2011interplay,kupavskii2012prediction,tsur2012s,cheng2014can}. These properties may include the size of the network or a subgraph (e.g., size of a community), the influence of cascades or contagions in the network (e.g., number of adopters), metrics of individual nodes or structures (e.g., degree or diameter), or even external properties that are not directly observed from the network structure (e.g., prestige, productivity or revenue of a node or a community). All these properties change over time, and their dynamics can be generally referred to as the \textit{growth} of a network\footnote{The \textit{growth}  refers to both the increment and decrement of the dynamic properties of the networks, i.e., positive or negative \textit{growth}.}. Indeed, the prestige of an individual node grows with the size of its ego-network. The influence range of a diffusion grows with the size of the diffusion network, subgraph of people who have adopted the diffusion. Accurate prediction of network growth has many valuable applications. For example, predicting the growth of research communities helps scientists to identify promising research directions; predicting the growth of social groups helps social network vendors optimize their marketing strategies; predicting the growth of the diffusion of a rumor helps analysts to estimate its potential damage and apply intervention in time. 

%The majority of prior work design a set of features, typically based on text content, network structures, or some heuristics, to predict either the growth of a community~\cite{backstrom2006group,romero2011interplay}, or the influence of a cascade~\cite{kupavskii2012prediction,tsur2012s,cheng2014can}. The commonality of these studies is that they are trying to predict a network property, which could vary overtime. We in general refer to this property as \textit{growth} hereafter.

Taking a typical data mining perspective, most existing methods extract features from both the network itself and any external information sources available. A function is learned that takes these features as input and outputs a predicted value of the network property in the future \cite{backstrom2006group}. From many explorations on different genres of networks, there has been a consensus in literature that features extracted from the topological structure of the network (a.k.a., the \textit{graph}) are generally very informative in these prediction tasks~\cite{backstrom2006group,cheng2014can}. As a comparison, other types of information, e.g., content or demographics, are only useful in certain scenarios. For example, the content of a hashtag is predictive to its diffusion \cite{yang2012we} and homophily (e.g., similar demongraphics) is predictive to the growth of social groups \cite{chen2015does}, but these effects are not generalizable to other networks and other dynamic properties. In this study, we focus on investigating the predictive power of the \textit{graph structure} of a network on its growth. 

Existing structural features are typically hand-crafted based on theoretical and empirical findings in the social network literature. For example, open triads with two strong ties are likely to be closed in the near future \cite{easley2010networks}; dense communities are resistant to novel information and they grow slower than others \cite{guimera2005team}; nodes spanning structural holes are likely to gain social capital and experience a rapid growth of its prestige and other properties \cite{burt2000network}. Features such as network density, clustering coefficients, triadic profiles, and structural holes are therefore designed to implement these intuitions and represent the graph structure. 

%However, previously adopted features are typically simple or hand-crafted, leading to potential loss of information. Therefore we want to ask: to what extent can we predict growth of information networks purely by utilizing the power of network topology? This question is important in that there are many cases where only information of network structures is available. For example, Facebook users might not want to disclose their posts due to concerns of privacy; additional information attached to networks is few, or at least hard to come by, for networks like Internet routing networks and transportation networks.  Even though rich information is available, equipped with a powerful tool that could fully utilize structural information, we can anticipate a big boost of performance in prediction.

Despite the success in predicting network growth, there are observable issues of representing the topological structure of a network using these hand-crafted features. Some of them only describe a global property of the network, such as network density or degree distribution; some of them provide a fine-grained description of local structures but fail to capture global information, such as triads and other substructures; others lie between the two extremes, such as structural holes. None of these features is able to fully represent both the local and the global structure of a graph and the complex interaction between local and global properties.
%To answer the proposed question, it is necessary to address at least two issues. First, we need to find a way to represent the network. For some prediction scenarios, the skeleton of a network might be strongly indicative, while in other scenarios, the presence of some local structures, e.g., cliques, could be more informative. As a result, hand-crafted features could fall short, as they can easily lead to information loss. On the other end of the spectrum, an adjacency matrix could store the entire structural information. However, it causes problems in the second issue -- the representation of the network is not unique. Permutations of an adjacency matrix all map to one identical network. Simply feeding a set of adjacency matrices to a learner can hardly learn any useful information.
On the other hand, these heuristic features usually have a limited characterization power of networks, as many networks may share the same feature representation. For example, most real-world networks at scale may have a similar (power-law) degree distribution, and two very different networks may happen to have the same ratio of closed triangles. Taking a machine-learning point of view, we are intrigued by the following questions: \textit{what is a suitable representation of network structure and how effective is such a representation when used to predict network growth?}

Our answers to the two questions are inspired by the recent developments in deep learning and graph representation. We introduce a graph descriptor that is based on the Heat Kernel Signature (HKS) ~\cite{sun2009concise}, which serves as a universal low-level representation of the topological structures of networks. HKS has been successfully employed in representing the surface of 3D objects~\cite{fang20153d,xie2015deepshape}. By modeling the amount of heat flow over nodes of a network over time, HKS successfully stores both the global and the local structural information of the entire network. Using a histogram to describe the probability distribution of heat values at a series of time points~\cite{fang20153d,xie2015deepshape}, isomorphic networks (networks with the same topological structure) can be mapped to a unique representation at little loss of structural information. However, unlike 3D objects which are composed of polygon meshes, the structures of networks vary in shape, size, and complex local structures. To address this issue, some computations of HKS need to be approximated carefully. %This necessitates the reinvestigation as to whether we can still extract useful information for our prediction task based on HKS output.
Inspired by the semantics of the HKS-based graph descriptors, we propose a multicolumn, multiresolution neural network that learns latent hierarchical representations of graphs on top of the HKS-based graph descriptor. The proposed deep neural network, named DeepGraph, predicts network growth in an end-to-end process.

%As a large amount of information is preserved in the raw output of HKS, applying a simple regressor, e.g., linear regression, could fail to fully take advantage of the power of HKS. Therefore, we employ deep learning techniques to learn hierarchically latent representation from the network.

%In this work, we resolve the problem of predicting network growth purely from the perspective of network topology. We introduce a new way, HKS, to represent the information networks, and use deep learning techniques to learn latent representation of networks in an end-to-end manner. 
We conduct extensive experiments to evaluate the effectiveness of DeepGraph. Different growing properties are predicted for five genres of real-world networks, including cascade networks and ego-networks.  Empirical results show that our method outperforms baseline approaches that use alternative graph representations, hand-crafted features, or existing deep learning architectures. High-level representations learned by DeepGraph well connect to existing findings in the social network literature.

The rest of the paper starts with Section~\ref{sec:related}, which summarizes the related literature. In Section~\ref{sec:method}, we formulate the data mining problem and describe how HKS is used to represent the network structure and how the deep neural network is architected. We present the design and the results of empirical experiments in Section~\ref{sec:exp_setup} and Section~\ref{sec:exp}, and then conclude the paper in Section~\ref{sec:conclusions}.

\section{Related work}
\label{sec:related}
%Our study focuses on finding a suitable representation of the topological structure of a network and using this representation to predict network growth through deep neural networks. It is related to the following lines of work in litearture. 

Predicting the growth of networks or the evolution of certain properties of networks has been widely studied. People attempt to predict the dynamics of various network metrics or aggregated activities in a network, e.g., the number of up-votes on Digg stories~\cite{szabo2010predicting}, the number of newly infected nodes in diffusion~\cite{yang2010modeling}, the growth of a community~\cite{backstrom2006group,romero2011interplay}, or the dynamics of a cascade~\cite{kupavskii2012prediction,tsur2012s,cheng2014can}. In these studies, a set of problem-specific features are usually manually designed based on the network structure, textual content, user demographics, historical statistics, and other sources of information. Among them, the features extracted from the network structure are both effective in individual tasks and robust across different tasks. In this work, we limit our focus on information purely from the network structure.
%, hoping the results can be generalized to many network prediction tasks.

Finding a suitable representation of the topological structure of a network has always been a critical preliminary step of network analysis. Conventionally, a network is represented as an adjacency matrix or a sparse list of edges. However, these lossless representations do not effectively present the structural characteristics of the network. Moreover, they are sensitive to the manipulation of node orders, making networks with the same topological structure mapped to different representations. Other approaches represent the network structure with a series of network metrics and/or a set of structural patterns (e.g., triads~\cite{kossinets2006empirical}, quads~\cite{ugander2012structural}, or meta-paths~\cite{sun2011pathsim}). Arbitrary higher-order substructures can be included, such as communities and structural holes.
%Compared to the adjacent matrix and list of edge representations, 
These bag-of-substructures better capture local patterns of the network structure.The major problem of this approach is that it is computationally infeasible to enumerate high-order substructures, and low-level substructures have limited representation power of the global structure of the network. As a result, many different networks may share the same or similar bag-of-substructures.
% However, bag-of-substructures also result in a considerable loss of information, and different network structures may be easily mapped to the same set of substructures. 

\begin{comment}
An alternative representation of a graph structure is a ``bag-of-substructures,'' analogical to the ``bag-of-words'' representation for text documents. Specifically, every graph can be represented as a set or a count vector of structural patterns, such as nodes, edges, and triads. Arbitrary higher-order substructures can be included, such as communities and structural holes. The major problem of this approach is that it is computationally infeasible to enumerate high-order substructures, and low-level substructures have limited representation power of the global structure of the network. As a result, many different networks may share the same or similar ``bag-of-substructures.''
\end{comment}

In graph classification, a myriad of graph kernel methods are proposed which compute pairwise similarities between graphs ~\cite{kashima2004kernels,bai2015quantum,shervashidze2011weisfeiler,shervashidze2009efficient}. For example, \textit{graphlets}~\cite{shervashidze2009efficient,ugander2013subgraph} computes the graph similarity based on the distribution of induced, non-isomorphic sub-graphs.
%of size $k$. 
%The similarity matrix can then be fed to a classifier. 
Some other graph kernels integrate frequent graph mining into the model training process~\cite{saigo2009gboost,ranu2009graphsig}. Graph kernels provide an indirect representation of networks so that similar structured networks yield a high value through the graph kernel function. The burden of graph kernels is the design of effective kernels. In the paper, we compare existing graph kernels to highlight the flexibility of our model.
%Graph kernels are compared with our model systematically. 

Recently, researchers have started to apply deep learning to network structure representation learning. Several proposals have been made to learn a low-dimensional vector representation of individual nodes by considering their neighborhood~\cite{tang2015pte,perozzi2014deepwalk,grovernode2vec}. Deep learning techniques have also improved graph kernels for graph structure learning\cite{yanardag2015deep,yanardag2015structural,subgraph2vec}. Recently, Niepert et al.~~\cite{niepert2016learning} applied convolution over receptive fields constructed by sequence of neighboring nodes. These methods focus only on the local structure of a graph and graph kernels require expensive pairwise comparisons.
%, leading to loss of information of the global structure. 
% Furthermore, expensive pairwise comparisons are needed construct graph kernels. %, making them hard to be applied to real-world prediction tasks. We propose to represent the network structure with a heat kernel signature based graph descriptor, which preserves both the local and the global patterns of the network, provides a near-unique mapping of isomorphic networks, and is yet efficient to compute. 
In the paper, we compare our model to these alternative deep learning approaches and show the performance advantage of our model. 

%The use of heat kernel signature is naturally related to the application of heat kernels. 
\textit{Heat kernels} have been studied for the task of graph clustering~\cite{bai2004heat}, graph partitioning~\cite{fang2015heat}, and modeling social network marketing processes~\cite{ma2008mining}. These applications rely on the raw output of heat kernels for a variety of tasks, rather than developing a \textit{signature}, nor do they abstract representation of graphs base on heat kernels. In the community of computer vision, Heat kernel signature has been successfully used to model 3D objects~\cite{sun2009concise,fang20153d,xie2015deepshape}, whose surfaces are defined by polygon meshes, a network composed of simple convex polygons. In contrast, real-world networks are consist of various shapes, sizes, and local structures. How to represent arbitary networks with heat kernel signatures and how to predict network growth using such a signature remain a challenging question to be studied.

%input{tex/ProblemDef}
\renewcommand{\vec}[1]{\mathbf{#1}}

\section{DeepGraph for Network\\ Growth Prediction}
\label{sec:method}

%label{sec:problem_def}
% TODO: high level description of pipeline, each component in turn
We propose a unified predictive neural network model to learn graph structure representation for network growth prediction problem. 
%An example of our model is shown in Figure~\ref{fig:pipeline}~(b). 
The proposed predictive model, named DeepGraph, combines heat kernel signature and deep neural networks. Below we describe the two key components of our model, (1) a heat kernel signature based graph descriptor and (2) a deep multi-column, multi-resolution convolutional neural network, in turn, following a brief definition of the network growth prediction problem.

%In this section, we introducing the basic notations and concepts, followed by defining the task of network growth prediction. 

\subsection{Problem Formulation and Notations}
%\paragraph*{\bf Network notations} 
Given a real-world network snapshot at time $t$, denote
its graph structure as $\mathcal{G}^{(t)}=(V,E)$, with a set of nodes $V$ and a set of edges $E$. A node $i\in V$ represents an entity (e.g., an actor in a social network or a paper in a citation network), an edge $(i,j)\in E$ represents a relationship (e.g., friendship, citation, or influence) between node $i$ and node $j$. 
An adjacency matrix $\vec{W} \in \mathds{R}^{|V|\times|V|}$ encodes the topological structure of the graph $\mathcal{G}$. In this work, we consider the binary adjacency matrix. Its element $w_{ij}$ is 1 if and only if $(i,j)\in E$ and 0 otherwise. 

A network property is a function that maps a graph structure $\mathcal{G}^{(t)}$ to a property value $y^{(t)} \in \mathds{R}$. For example, a network property could be the number of friends given a user's Facebook ego-network. A network growth predictor is a function that maps a graph structure $\mathcal{G}^{(t)}$ to a property value $y^{(t')}$ at time $t'$, satisfying $t'>t$. For example, a network growth predictor could map a user's Facebook ego-network of this year to the number of friends next year.

The network growth prediction can be naturally formulated as a supervised learning problem. Specifically, the problem is to derive a network growth predictor $f$ given a training set of tuples $\{(\mathcal{G}^{(t_{i})}_{i}, y^{(t'_{i})}_{i})\}_{i=1}^{M}$ to minimize the prediction error over a test set of tuples $\{(\mathcal{G}^{(t_{j})}_{j}, y^{(t'_{j})}_{j})\}_{j=1}^{N}$ satisfying $\forall i$ $t'_{i} > t_{i}$, $\forall j$ $t'_{j} > t_{j}$, $\min_j(t_{j})$ $>$ $\max_i(t_{i})$, and $\min_j(t'_{j})$ $>$  $\max_i(t'_{i})$. The time ordering constraints highlights the practical motivation that we are interested in using historical data to predict future properties of current networks.\footnote{In practice, researchers focus on a specially case of the network growth prediction problem with the equal interval increment constraint, $t'_{j}-t_{j}=t'_{i}-t_{i}=C>0$~\cite{kupavskii2012prediction,tsur2012s}.}  To apply a machine learning algorithm, it is critical to first represent the graph $\mathcal{G}^{(t)}$ computationally, such as using a vector of features.

%that we are interested in predicting Specifically, we focus on supervised learning approach to solve the network growth prediction problem. 
%The network growth prediction problem can be naturally solved through a supervised learning approach if we observe many such timestamped tuples $\{(G^{(t)}, y^{(t')})\}$. A machine learning algorithm, such as a linear regression, can be applied using these tuples and make predictions for graphs whose $y^{(t_1)}$ are not observed. To do that, it is critical to first represent the graph $\mathcal{G}^{(t)}$ computationally, such as using a vector of features.
% TODO: add a discuss about train data time and test data time

Notice that such a representation can consider different sources of information when a particular type of real-world network is considered: the graph structure, the profiles of the nodes, the semantics of the edges, the activities of the nodes, and the information content being circled within the network. We focus our investigation onto solely the graph (topological) structure, as the structural information is ubiquitously available in all types of networks and it is known to be predictive to the growth of the network. We leave a multi-modal representation of the network for future work. 

\subsection{Heat Kernel Signature based\\ Graph Descriptors}
% \subsubsection{The HKS Graph Descriptor}
The motivation in adopting Heat Kernel Signature (HKS) is its theoretical proven properties in representing graphs: HKS is an \textbf{intrinsic} and \textbf{informative} representation for graphs~\cite{sun2009concise}. \textbf{Intrinsicness} means that isomorphic graphs map to the same HKS representation, and \textbf{informativeness} means if two graphs have the same HKS representation, then they must be isomorphic graphs. Our HKS-based graph descriptor builds on the theoretical properties of HKS and further provides universal representations for graph with different sizes in network growth prediction.
%for network growth prediction. We provide details of our HKS based graph descriptor.

%In order to approach the task of predicting network growth, we need to tackle two problems for the network representation: (1) it could preserve global to local information, and (2) the representation is unique for each network. In this paper, we investigate the possibility of utilizing Heat Kernel Signature (HKS)~\cite{sun2009concise} to describe a network. By modeling the flow of heat over nodes in the network at different diffusion time points, HKS could store both global and local structural information of the entire network. When histogram is used to estimate the probability distribution of HKS values~\cite{fang20153d,xie2015deepshape}, a network can be invariantly mapped to a unique representation.

%The overall pipeline to process network data, and generate prediction for network growth is shown in Figure~\ref{fig:pipeline}. In the following subsections, we first introduce HKS, together with discussion of adapting it to the case of information networks. Next, we present our structure of deep neural network used in this task.

%\subsection{Heat Kernel Signature (HKS)}
\paragraph*{\bf Heat kernel function}
%Assuming that nodes of a graph are embedded in a manifold, where edges between nodes reflect the geometry of this manifold. In order to describe this manifold, we can suppose initially there is a unit amount of heat at any node $i$, which will flow along edges overtime. The edge weight decides the speed of diffusion on edges, while the number of edges incident to a node (or more precisely the sum of weight) decides the speed on nodes. If we record the amount of heat flow for any pair of nodes overtime, the structure of the network could be well preserved. This forms the intuition of heat kernel.

Formally, the heat kernel $h_z(i,j)$, a function of two nodes $i$, $j$ at any given diffusion step $z$, denotes the amount of aggregated heat flow through all edges among two nodes after diffusion step $z$\footnote{The diffusion is simulated for a given graph snapshot. The heat kernel computation does not require graph snapshot at other timestamps. The diffusion step $z$ should not be confused with the network timestamp $t$.}. In computer vision, graphs are stored as meshed networks and heat kernels are computed by finding eigenfunctions of the Laplace-Beltrami operator~\cite{sun2009concise}. However, meshed networks are not available for most real-world networks. Instead, we use eigenfunction expansion of a graph Laplacian~\cite{sun2009concise,bai2004heat} to compute the heat kernel for information networks. 
%in the case of information networks, the heat kernel computation requires a normalized graph need to be approximated. Some work use eigenfunction expansion of Laplacian~\cite{sun2009concise,bai2004heat} for this purpose. 
Given a graph $\mathcal{G}=(V,E,W)$, the graph Laplacian is defined as:
\begin{equation}
\vec{L} = \vec{D}-\vec{W}
\end{equation}
where $D$ is a diagonal degree matrix with diagonal entries being the summation of rows of $W$: $D_{ii} = \sum_j w_{ij}$.
The normalized Laplacian of the graph is given by
\begin{equation}
\vec{L_N} = \vec{D}^{-\frac{1}{2}}\vec{L}\vec{D}^{-\frac{1}{2}}
\end{equation}

The heat kernel is then defined as
\begin{equation}
h_z(i,j) = \sum_{k=1}^{|V|} e^{-\lambda_k z}\phi_k(i)\phi_k(j)
\label{equ:heat_kernel}
\end{equation}
where $\lambda_k$ is the $k$-th eigenvalue of the normalized Laplacian $\vec{L_{N}}$ and $\phi_k$ is the $k$-th eigenfunction s.t. $\sum_i |\phi_k(i)|^2 = 1$. 
%The eigenvalues are ordered ascendingly by magnitude. 
Note that the eigenvalues might be unreal in the case of directed graphs. There has been studies on how to tackle this problem~\cite{chung2005laplacians}. In this work, for simplicity, we convert directed graphs to undirected ones by applying $\vec{W} = (\vec{W}+\vec{W}^{\intercal}) / 2$.

% The family of heat kernel functions $\{k_{z}(v,.)\}_{t>0}$ provides a point signature for $\forall v\in V$. However, the computation complexity of such point signature is overwhelming since the signature is defined on the product of temporal and spatial domain. Heat kernel signature provides an efficient computation by removing a large amount of redundant information in the naive point signatures. 

\paragraph*{\bf Heat kernel signature}  
Heat kernel signature was introduced to mitigate the computation bottleneck of using heat kernel functions in representing graphs. Both heat kernel and heat kernel signature are proven to be intrinsic and stable against noises. However, the computation complexity of using heat kernel as a point signature is overwhelming since the point signature, $\{k_{t}(v,.)\}_{t>0}$ ,  is defined on the product of temporal and spatial domain.  Heat kernel signature simplifies the computation by considering only a subset of product of temporal and spatial domain while keeping as much information as possible. 
Specifically, heat kernel signature reduces the computation complexity by only requiring $h_{z}(v,v)$ over a finite set of $N$ diffusion steps $z\in\{z_{1},z_{2},...,z_{N}\}$ for $\forall v \in V$ without losing the intrinsic and informative properties. 

%Heat kernels are intrinsic and informative representation of graphs. These properties make the heat kernel a very lucrative candidate for a point signature. 
Formally, a heat kernel signature (HKS) is a matrix $\vec{H} \in \mathds{R}^{|V|\times N}$ satisfying
\begin{equation}
H_{ij} = h_{z_{j}}(i,i)
\label{equ:HKS}
\end{equation}
%where $t_n$ denotes the diffusion time of the $n$-th sample point. 
These time points are sampled with equal difference after logarithm~\cite{sun2009concise}, such that $\log z_{n} - \log z_{n-1} = \log z_{n+1} - \log z_{n}$.

%{\color{red} Add a paragraph about the property/benefit of HKS: why is it a good graph representation; why we don't use heat kernel directly?} 

%\subsubsection{Graph Descriptor}
%\label{sec:HKS_descriptor}
\paragraph*{\bf Graph descriptor}
\begin{figure}[h!]
\begin{minipage}[t]{0.5\textwidth}
\centering
\begin{minipage}[t]{0.24\textwidth}
\centering
\includegraphics[width=0.945\textwidth]{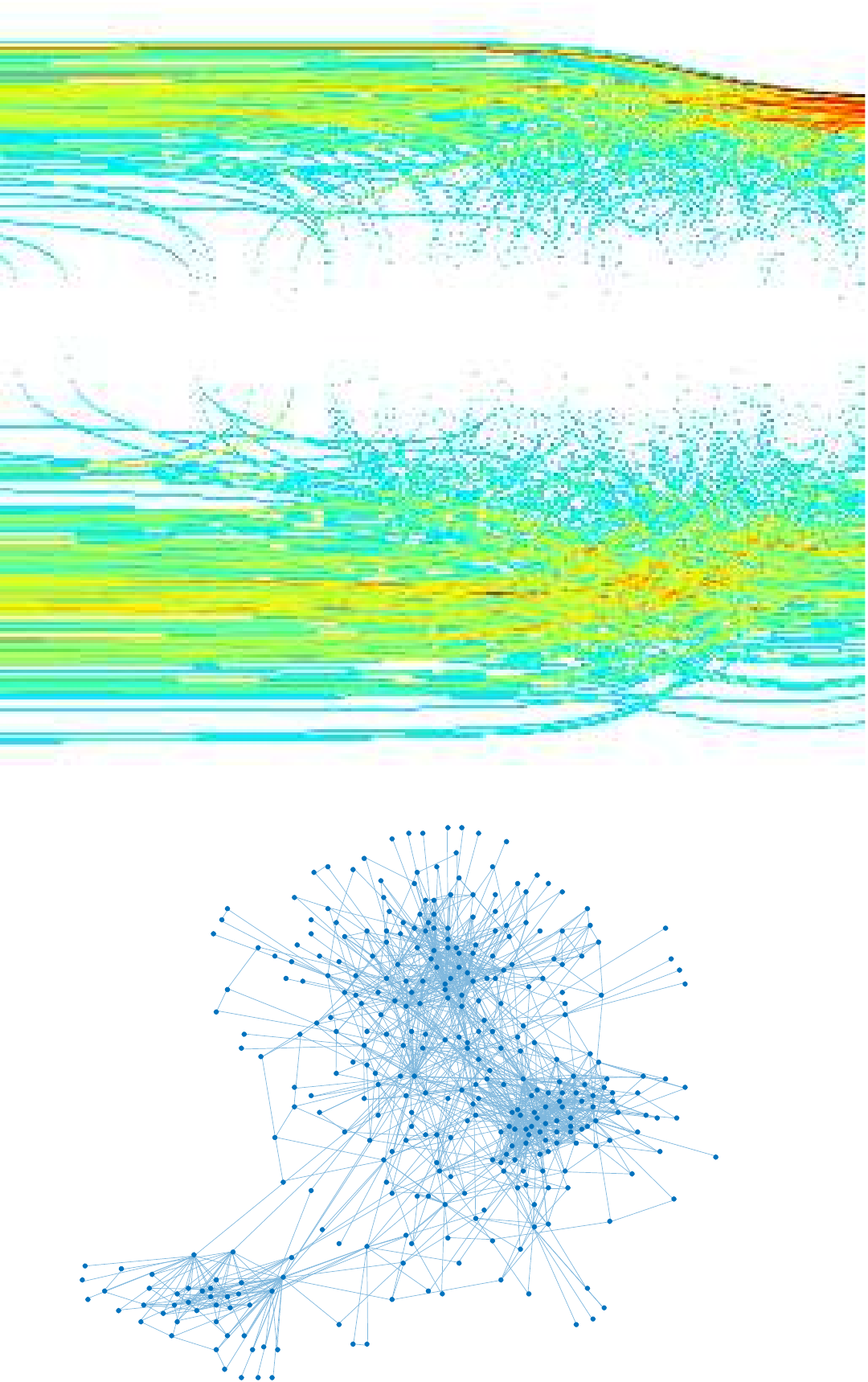}\\
(a)
\end{minipage}
\begin{minipage}[t]{0.24\textwidth}
\centering
\includegraphics[width=0.98\textwidth]{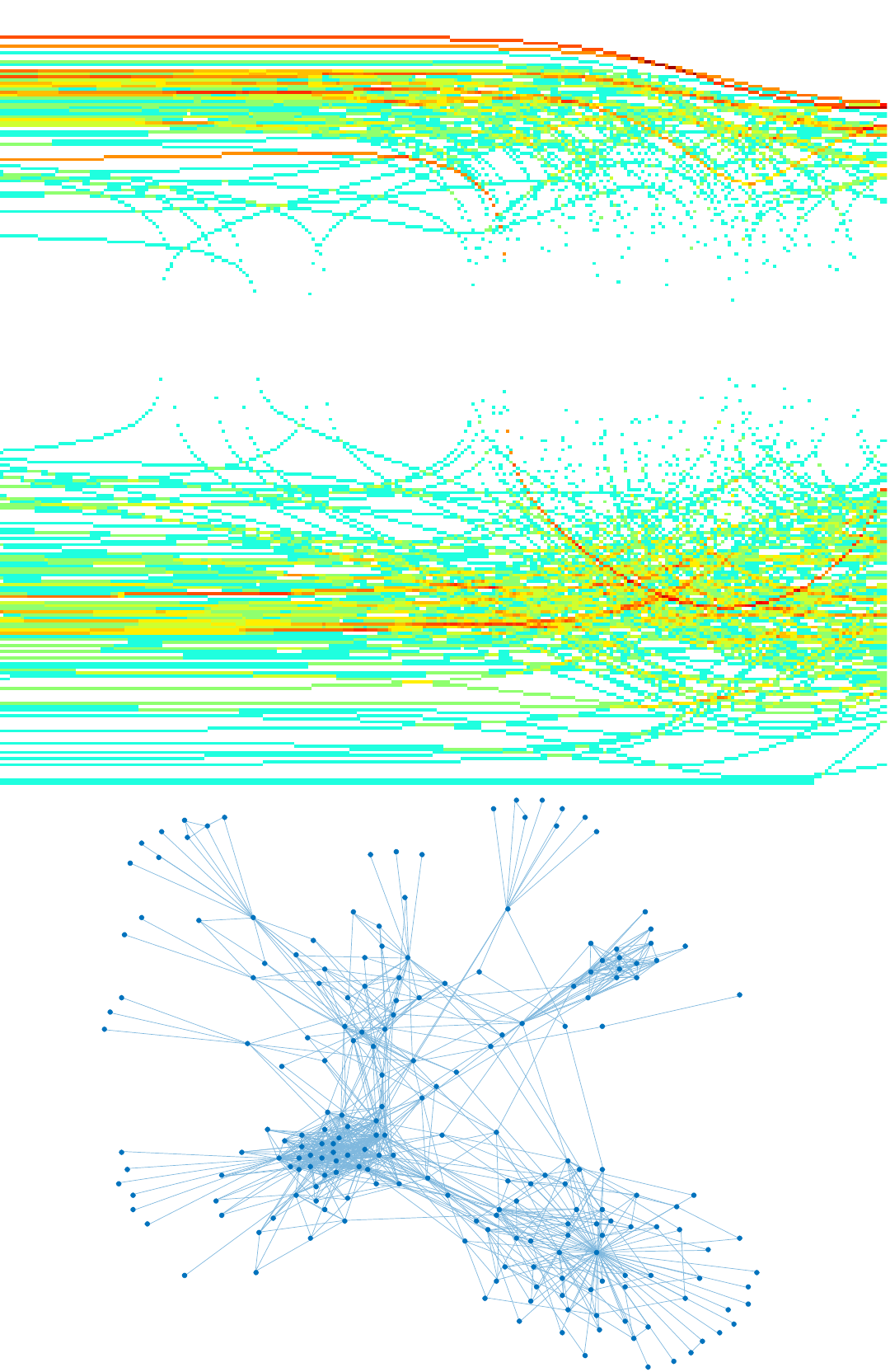}\\
(b)
\end{minipage}
\begin{minipage}[t]{0.24\textwidth}
\centering
\includegraphics[width=0.945\textwidth]{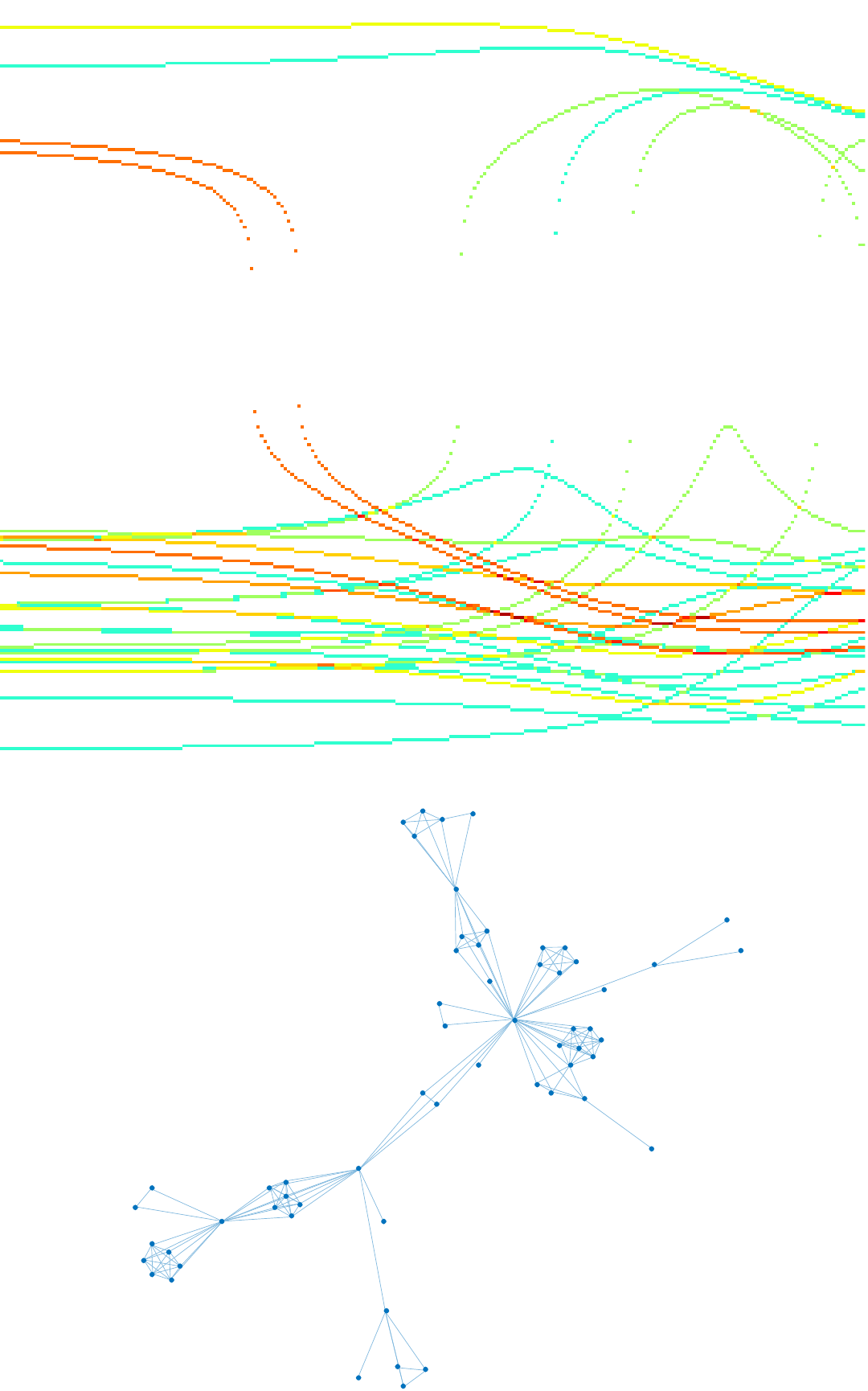}\\
(c)
\end{minipage}
\begin{minipage}[t]{0.24\textwidth}
\centering
\includegraphics[width=0.945\textwidth]{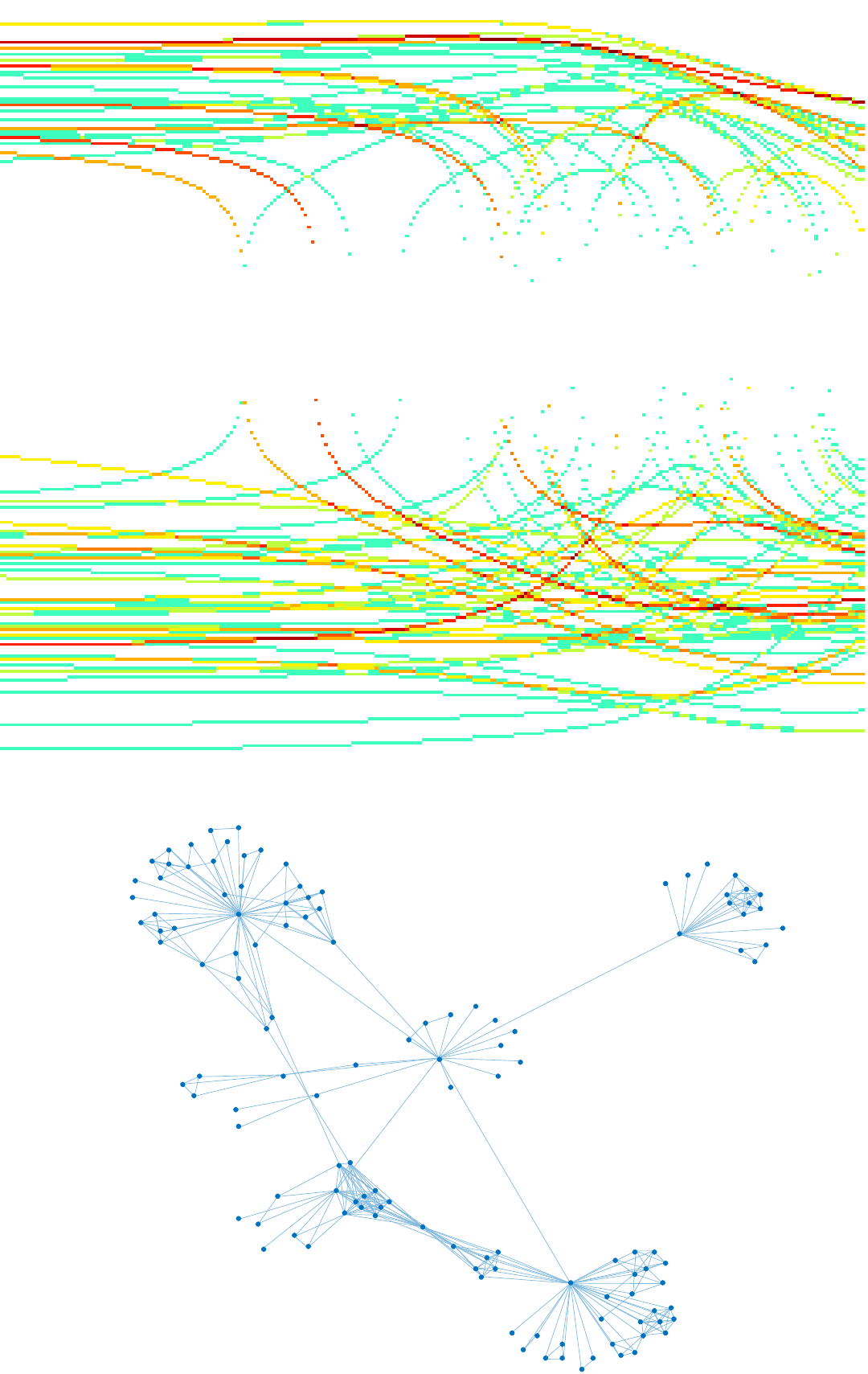}\\
(d)
\end{minipage}
\caption{Examples of HKS-based graph descriptors. The first row shows our graph descriptors for graphs in the second row. Figure (a) and (b) are subnetworks from Facebook~\cite{viswanath-2009-activity}. Figure (c) and (d) are some authors' collaboration networks built from ACL Anthology~\cite{Radeval.09}.}
\label{fig:hks_examples}
\end{minipage}
\end{figure}

The practical issues in combining HKS and deep neural networks are that we need a global vertex indexing to guarantee the uniqueness and that the size depends on $|V|$. We further process heat kernel signature $\vec{H}$ into a universal representation independent of $|V|$ using a histogram conversion. Specifically, we use histograms to estimate the distribution of HKS values in each column\footnote{The bin ranges are aligned column-wise on the training data.}. By denoting $N_{B}$ the number of bins used in the histogram, we obtain a universal descriptor $\vec{S} \in \mathds{R}^{N_{B}\times N}$. Unlike HKS, the new descriptor is independent of vertex ordering and vertex number. We call this final matrix \textit{graph descriptor}, $\vec{S}(\mathcal{G})$, as it is adapted to describe information networks. Figure~\ref{fig:hks_examples} shows four examples of our graph descriptors for real world graph structures. 

\paragraph*{\bf Graph descriptor vs. adjacency matrix} We have described the process in converting an adjacency matrix into our graph descriptor, which is then passed through a deep neural network for further feature extraction. All computation in this process is to obtain a more effective low-level representation of the topological structure information than the original adjacency matrix.

First, isometric graphs  could be represented by many different adjacency matrices, while our graph descriptor would provide a unique representation for those isometric graphs. The unique representation simplifies the neural network structures for network growth prediction.

Second, our graph descriptor provides similar representations for graphs with similar structures. The similarity of graphs is less preserved in adjacency matrix representation. Such information loss could cause great burden for deep neural networks in growth prediction tasks. 

Third, our graph descriptor is a universal graph structure representation which does not depend on vertex ordering or the number of vertexes, while the adjacency matrix is not. 

\paragraph*{\bf Time complexity} The major overhead of computing graph descriptors lies in the calculation of eigenvectors. The time complexity of computing eigenvectors is $\mathcal{O}(K|V|^{2})$ where $K$ is the number of eigenvectors. Our graph descriptors finish in acceptable time frame for real world network data. The data description and time complexity analysis are in Section~\ref{sec:exp_setup}. 

\paragraph*{\bf Semantics of graph descriptor} The rows and columns in our graph descriptor reflect the network topology from different perspectives. The rows express the heat density dynamics over diffusion steps, and the columns capture the static heat density patterns for a given diffusion step. Successive rows or columns express higher-order properties of the topology structure information. Such representational properties motivate the adoption of row-wise and column-wise convolution networks for feature learning.  
%by computing only the first $K$ eigenvectors and eigenvalues. In our experiments, we found that this approximation can already yield very positive results. We consider this a significant advantage compared to ``bag-of-substructures'' representations, as counting higher-order substructures (e.g., quads) is very costly. 

\subsection{Deep Graph Descriptor}
\begin{figure*}[ht]
\centering

\includegraphics[width=0.98\textwidth]{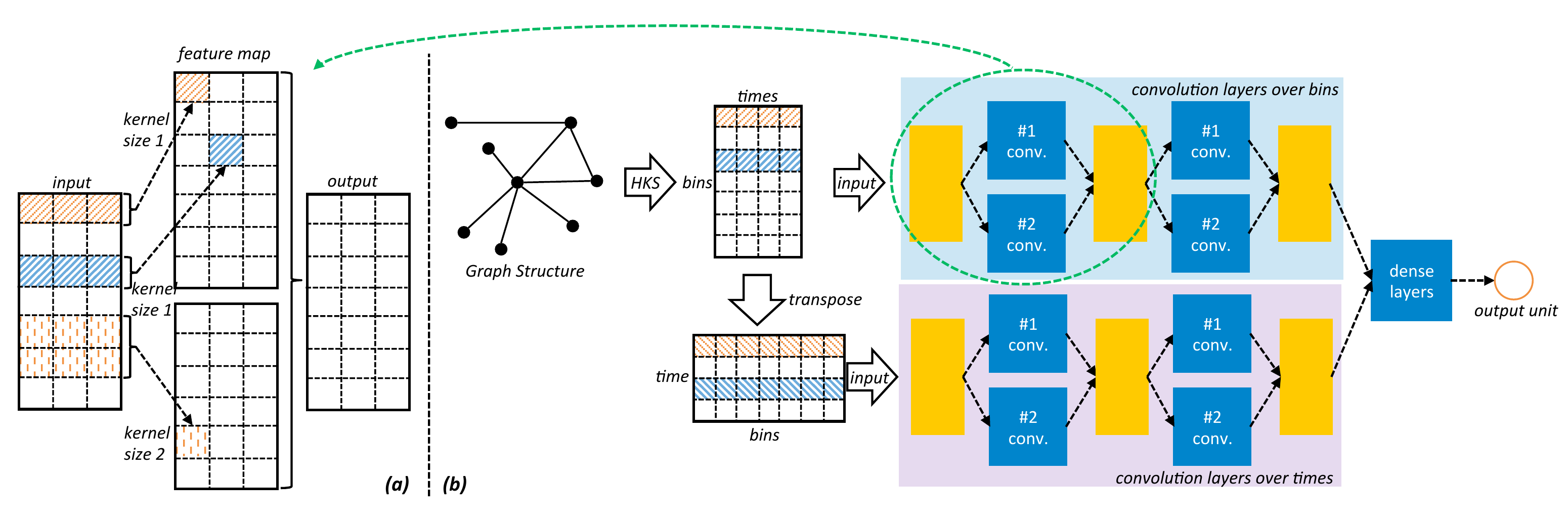}
\caption{(a) An example of the multiresolution convolution unit with two kernel sizes. (b) An example of the multicolumn, multiresolution deep neural network model for network growth prediction with two convolution layers. }
\label{fig:pipeline}
\end{figure*}

As information abounds in the raw representation extracted by the HKS-based graph descriptor, applying a simple regressor, e.g., linear regression, could fail to fully extract useful information from it. In contrast, deep neural networks (DNN) have achieved tremendous success in learning latent representations from raw inputs in a compositional hierarchy. Combining DNN and HKS-based graph descriptor together thus offers an opportunity to address the graph structure representation challenges in predicting network growth. Inspired by the semantics of the graph descriptors, we propose a deep multicolumn, multiresolution convolutional neural networks for the network growth prediction task.  

% \item HKS is universal for different numbers of vertexes $|V|$ while the adjacent matrix is of size $|V|\times|V|$ by definition. 

% At last, either rows or columns in the adjacent matrix carry only local connectivity information while rows and columns in HKS present both global and local graph structure information. Such property of HKS inspires the adaption of 1-D convolution. 

% In summary, HKS is a more effective and nature representation for deep neural network compared to adjacency matrix representation.

\paragraph*{\bf Multiresolution convolutions} Our model builds on the multiresolution 1-D convolution (MrConv) which maps an input matrix into a feature map matrix.  Specifically, let $\vec{x}_{i} \in \mathds{R}^{k}$ denote the $i$-th row of the input matrix. The input is then represented as $\vec{x}_{1:n}=\oplus_{i=1}^{n}\vec{x}_{i}$ where $\oplus$ is the concatenation operator and $n$ is the number of rows. The 1-D convolution with a filter size $m$ apply a filter  $\vec{w}$ $\in$ $\mathds{R}^{mk}$ to each possible window of $m$ rows to produce a new feature vector $\vec{c} = [c_{1}, c_{2}, ..., c_{{n-m+1}}]$. The feature $c_{i}$ is generated from a window of $m$ rows $\vec{x}_{i:i+m-1}$ by:
\begin{equation}
c_{i}=g(\vec{w}^{\intercal}\vec{x}_{i:{i+m-1}}+b)
\end{equation}
where $b$ $\in$ $\mathds{R}$ is a bias term and $g$ is a non-linear function such as a hyperbolic tangent function or a rectified non-linearity function.

We have described the process by which \textit{one} feature vector $\vec{c}$ is extracted from \textit{one} filter. Our multiresolution convolution (MrConv) layer uses multiple filters with varying filter sizes to obtain multiple resolution features. Specifically, one MrConv layer has $l$ different convolution filter sizes $\{m_{1}, m_{2}, ..., m_{l}\}$. The filter of size $m$ generates a corresponding feature vector $\vec{c}^{(m)}$. Feature vectors generated by different filter sizes are then concatenated into one vector $\vec{c}^{*}=\oplus_{i=1}^{l}\vec{c}^{(m_{i})}$. Moreover, we extend each filter size to have $d$ different filters. The final output feature map is a matrix $\vec{O}$ where each column is a feature vector $\vec{c}^{*}$ and there are $d$ columns. 
\begin{equation}
\vec{O}=\Big( \vec{c}^{*1},\vec{c}^{*2},...,\vec{c}^{*d}  \Big)
\end{equation}
\begin{comment}
The $d$ feature vectors generated from convolution kernel of size $h_{j}$ form a feature map with $n$$-$$h_{j}$$+$$1$ rows and $d$ columns, where the $i$-th column is the feature  vector from the $i$-th filter. Then feature maps from different kernel sizes are then concatenated in rows to form a final feature map as output for next multiresolution convolution layer. Multiple multiresolution convolution layers are then stacked to form a deep multiresolution neural network. %with $(n+1)m-\sum_{i}h_{i}$ rows and $d$ columns.  TODO: a feature map?
\end{comment}

An example of our MrConv is shown in Figure~\ref{fig:pipeline}(a). The example MrConv layer has two different filter sizes $\{1,2\}$. Each filter size has three different filters, whose feature vectors form different columns in the final feature map. Multiple multiresolution convolution layers are stacked to form our model. 

\paragraph*{\bf Multicolumn model} Inspired by the different semantics of rows and columns in the HKS-based graph descriptor, our model deploys a two network-column structure, as shown in Figure~\ref{fig:pipeline}(b). One column uses multiresolution 1-D convolution (MrConv) operations over the graph descriptor bins and the other one uses MrConv over diffusion times. The two columns extract different features from the graph descriptors at multiple resolution scales. Intuitively, the first column extracts statistical features of the density dynamics in diffusion.  The second column extracts features on static density pattern for different diffusion steps. Both kinds of features reflect the topology of the underlying graph structure, but explain the structure topology from different perspectives. A single column convolutional neural network can hardly extract such two kinds of features successfully. 

The feature maps from the two columns are then concatenated and passed through multiple dense (i.e. fully-connected) layers with non-linear activation functions. The output from the multiple dense layers are then passed through a final linear fully-connected layer with only one output unit. The output unit $\hat{y}$ is thus the network growth prediction of our model.   

\subsection{End-to-End Training}
Let $\textrm{McMrConv}(., \theta)$ denote the multicolumn multiresolution convolutional neural network with parameters $\theta$. The final output of our neural network given a graph $\mathcal{G}_{k}$ is represented as:
\begin{equation}
\hat{y}_{k}=\textrm{McMrConv}(\vec{S}(\mathcal{G}_{k}), \theta)
\end{equation}
Given a training data set $\{(\mathcal{G}_{k},y_{k})\}_{k=1}^{K}$, the deep neural network is trained to minimize the average squared error.
\begin{equation}
\mathcal{L}(\theta)=\frac{1}{K}\sum_{k=1}^{K}\Big( \textrm{McMrConv}(\vec{S}(\mathcal{G}_{k}), \theta)-y_{k}\Big)^{2}
\end{equation}
The HKS-based graph descriptor and the deep neural network assembles DeepGraph, an end-to-end deep architecture to predict network growth based on graph structure. % In the following sections, we evaluate the effectiveness of DeepGraph with empirical experiments.  

\section{Experiment setup}
\label{sec:exp_setup}
We compare our model with existing approaches, including hand-crated feature based linear or nonlinear regression, graph kernels and alternative deep learning approaches, on the network growth prediction problem. We then evaluate variants of our model to assign the credit of the  two key components, the HKS-based graph descriptor and the deep neural network.
\begin{comment}
We present empirical experiments that systematically evaluate the effectiveness of DeepGraph. The section describes the setup of these experiments, including data sets, evaluation metric, baselines for comparison, and parameter configurations.
\end{comment}

\subsection{Data sets}
When selecting real-world data sets for evaluation, we consider both popularity and diversity of the application scenarios. The five data sets we choose include social networks, scientific collaboration networks, information diffusion networks, and entertainment networks. Note that to train and test the network growth prediction algorithm, it is desirable to obtain a large number of time-variant networks, which is not directly available (e.g., there is one global Facebook network). So following~\cite{yanardag2015deep}, from each large network we extract the ego-networks (subgraph consisting of the neighbors of an ``ego'' node) of many individual nodes. The statistics of these data sets are presented in Table~\ref{tab:datasets}. Please note that due to the diverse nature of the data sets and the various precision of timestamps available, it is hard to apply an unified time frame for all data sets. Viewed from another perspective, this helps us evaluate the flexibility and generality of our methods, verifying whether it can be applied to any length and granularity of time frames. 

\begin{table}[h!]
\caption{Statistics of the data sets.}
\small
\begin{center}
\setlength{\tabcolsep}{3pt}
\begin{tabular}{c|c|c|c|c|c|c} 
\hline 
 & Dataset & Facebook & YouTube & AAN & IMDB & Weibo \\
\hline 
 & train & 12990 & 15258 & 8426 & 12500 & 14614 \\
\# graphs & val & 890 & 1283 & 713 & 1017 & 1092 \\
 & test & 2092 & 3273 & 1722 & 2407 & 2555 \\
\hline 
 & train & 399.9 & 147.6 & 271.4 & 197.3 & 71.1 \\
Avg. nodes & val & 397.5 & 167.3 & 302.4 & 208 & 76.6 \\
 & test & 436 & 165.8 & 402.4 & 216 & 64.9 \\
\hline 
 & train & 6800.8 & 1439.2 & 2079.8 & 7801.5 & 55.6 \\
Avg. edges & val & 6764.4 & 1626.1 & 2327.2 & 7847.7 & 95.8 \\
 & test & 7499.2 & 1620.9 & 3321.8 & 7964.5 & 100.1 \\
\hline 
 & train & 3.6 & 9 & 1.2 & 1.3 & 15.4 \\
Avg. growth & val & 3.8 & 10.8 & 1.2 & 1.3 & 16.7 \\
 & test & 3.4 & 9.3 & 1.2 & 1.3 & 14.1 \\
\hline 
 & train & 1.7 & 2.4 & 0.9 & 1 & 2 \\
Avg. scaled  & val & 1.8 & 2.2 & 0.9 & 1 & 1.8 \\
growth$^{1}$ & test & 1.6 & 2.1 & 0.9 & 0.9 & 1.9 \\
\hline 
 & train & 2007.6 & 2007.2 & 2009 & 2000 & 5h \\
Graph time$^{2}$ & val & 2007.7 & 2007.3 & 2010 & 2001 & 5h \\
 & test & 2007.8 & 2007.4 & 2011 & 2002 & 5h \\
\hline 
 & train & 2007.10 & 2007.4 & 2010 & 2001 & 3h \\
Growth time$^{3}$ & val & 2007.11 & 2007.5 & 2011 & 2002 & 3h \\
 & test & 2007.12 & 2007.6 & 2012 & 2003 & 3h \\
\hline 
k-hop ego-net$^{4}$ & all & 2 & 2 & 3 & 2 & - \\
\hline 
\end{tabular}
\flushleft\textit{1. Avg. scaled growth} scales label $y$ to $\log_2(y+1)$~\cite{kupavskii2012prediction,tsur2012s}.\\
\textit{2. Graph time} of 2007.6 means the graph is built by taking the snapshot of Jun. 1, 2007. For diffusions in Weibo, we use the first 5 hours to build graph, and the next 3 hours to compute growth.\\
\textit{3. Growth time} of 2007.10 means the growth is computed between its corresponding \textit{graph time} to Oct. 1, 2007. Graphs in train/val/test set do not overlap.\\
\textit{4. k-hop ego-net} for AAN is set to 3, due to its small size of 2-hop ego-nets. Weibo contains diffusion networks rather than ego-nets.
\end{center}
\label{tab:datasets}
\end{table}

%\begin{table}[h!]
%\caption{Data source and growth prediction setup.}
%\small
%\begin{center}
%\setlength{\tabcolsep}{3pt}
%\begin{tabular}{c|c|c|c} 
%\hline 
%source & graph node & graph edge & property to predict \\
%\hline 
%Facebook~\cite{viswanath-2009-activity} & user & friendship & the number of new friends \\
%\hline
%YouTube~\cite{mislove-2009-socialnetworksthesis} & user & friendship & the number of new friends \\
%\hline
%AAN~\cite{Radeval.09} & author & co-authorship & h-index \\
%\hline
%IMDB & actor/actress & co-stars & number of new movies \\
%\hline
%Weibo & user & following-followee & number of retweets \\
%\hline
%\end{tabular}
%\end{center}
%\label{tab:data}
%\end{table}
%The data sources and corresponding network growth prediction setup is shown in Table~\ref{tab:data}.
We follow the procedure described in~\cite{yanardag2015deep} to construct ego-nets. The Facebook data set is collected from the New Orleans networks~\cite{viswanath-2009-activity}, where nodes are Facebook users and edges are friendships. We derive the snapshot of ego-networks for each user according to the timestamps listed in Table~\ref{tab:datasets}, which is used to predict the number of new friends this user made in the next four months. %The snapshot of the ego-network by July 2007 is used to predict the number of new friends the ego-node made within the last 6 months of 2007.

As the YouTube~\cite{mislove-2009-socialnetworksthesis} data set also describes user friendships, it follows the same setting as Facebook. %We use the first three months of year 2007 to build ego-networks and to predict the number of new friends made in the fourth month of 2007.

The AAN data set~\cite{Radeval.09} is built upon scientific publications from the ACL Anthology\footnote{\url{http://aclweb.org/anthology/}}, where nodes are authors and edges are collaboration. Each author's ego-nets are extracted to predict her h-index in the next year.

IMDB is a movie co-star data set\footnote{\url{http://www.imdb.com/}}, where nodes are actors or actresses, and an edge is formed if they appear in the same movie. The ego-nets of each actor/actress is used to predict the number of new movies the actor/actress produced in the next year. %It would be ideal to use the average rating of movies casted by an actor/actress as the label. However, ratings of most movies are unavailable. As a result, we count the number of new movies in 2002 and 2003 as the label. We extend the prediction duration to 2 years, as many actors/actresses have no new movies in one year.

The Weibo data set~\cite{zhang2013social} contains a set of Sina Weibo users with their complete following-followee relationships, as well as 300,000 retweeting paths among these users. For an original tweet we construct its diffusion network, which is the subgraph (of the following-followee graph) with users who retweeted the tweet within the first 5 hours of the original tweet. We then predict the growth of this diffusion network in the next 3 hours (i.e., the number of new users who retweet this tweet). For simplicity, we ignore the direction of the edges and treat all graphs as undirected.

To examine whether we can truly predict future growth, we make sure of two important points: (1) the period to compute growth for test set is always later than that for training set; (2) one graph can only appear in one of the training, validation and test set. For Weibo data set, we sort all diffusion graphs by time of occurrence, and take the earliest 80 percent graphs for training, the next 5 percent for validation, and the last 15 percent for testing. For other data sets, for each node in the global network, they are randomly assigned to the training/validation/test set with probability of 0.8/0.05/0.15. Based on which set they are in, their ego-nets and growth are computed according to the time listed in Table~\ref{tab:datasets}. If a node has not yet been created for the given time, it is simply removed.

We notice that the growth of all the networks in general follows a power-law distribution, where a large number of networks did not grow at all. Therefore we downsampled 50\% graphs of each train/val/test set with zero growth (to the numbers shown in Table~\ref{tab:datasets}) and applied a logarithm transformation of the outcome variable (network growth), following~\cite{kupavskii2012prediction,tsur2012s}. The network growth are scaled logarithmically for two reasons. First, baseline methods with linear regression are sensitive to extremely large outcomes. Second, when a network grows to a considerably large scale,  we care more about its scale rather than the exact number.

Please note that our method is not limited to ego-networks -- data set Weibo is a set of diffusion graphs of tweets.

%We further randomly sampled all ego-nets to the numbers shown in Table~\ref{tab:datasets}. This sampling process is used to guarantee that all competing methods, especially those computationally expensive baselines, could finish in a reasonable amount of time. However, it is important to note that every data set contains at least 10,000 networks. From every data set, We randomly sample 90\% of the networks as the training set and use the rest for testing. %Partitioning the data into training and test set is a common practice for studies on deep learning~\cite{krizhevsky2012imagenet,lee2009convolutional}.

\subsection{Evaluation Metric}
We use mean squared error (MSE) as our evaluation metric, which is a common choice for regression tasks. Specifically, denote $\hat{y}$ a prediction value, and $y$ the ground truth value, the MSE is:
\begin{equation}
\textrm{MSE}=\frac{1}{n}\sum_{i=1}^n(\hat{y}_{i} - y_i)^2
\end{equation}
As noted before, $y$ in above equation is a scaled version of the original value $y^o$, that is $y=\log_2(y^o+1)$.

\subsection{Baseline methods}
We compare DeepGraph with methods from two categories: feature-based methods used for network prediction tasks, and alternative graph representation methods. 

\textbf{Feature based}. Many structural features have been designed for various network prediction tasks~\cite{backstrom2006group,romero2011interplay,tsur2012s,cheng2014can}. We select from them those that could be generalized across data sets, including:

\textit{Frequencies of k-node substructures} ($k\leq 4$)\cite{ugander2012structural}. This counts the number of nodes ($k=1$), edges ($k=2$), triads (e.g., the number of closed and open triangles) and quads.
%Most existing successes of network growth prediction are reported using ``bag-of-substructures'' to represent the graph structure. We follow this convention and represent a graph with a feature vector that consists of the frequencies of k-node substructures. For example, all types of 3-node substructures (a.k.a. triads) are displayed in Figure~\ref{fig:graphlets}. These features are directly used to train a linear regression model to predict the network growth.

\textit{Other network properties}: average degree, the length of the shortest path, edge density, the number of leaf nodes (nodes with degree 1), the number of leaf edges, the average closeness of all nodes, clustering coefficient, diameter, and the number of communities obtained by a community detection algorithm~\cite{blondel2008fast}.

\textbf{Graph kernels}. Following~\cite{niepert2016learning}, we compare with four state-of-the-art graph kernels: the shortest-path kernel (\textbf{SP})~\cite{borgwardt2005shortest}, the random walk kernel (\textbf{RW})~\cite{gartner2003graph}, the graphlet count kernel (\textbf{GK}) \cite{shervashidze2009efficient}, and the Weisfeiler-Lehman subtree kernel (\textbf{WL}) \cite{shervashidze2011weisfeiler}. In our experiment, the RW kernel does not finish after 10 days for a single data set, so we exclude it for comparison. This exclusion is also observed for the same reason in~\cite{niepert2016learning,yanardag2015structural}.

\textbf{-linear} and \textbf{-deep}. Feature based methods and graph kernels are usually trained on SVMs. We report linear regression instead, as SVM empirically generates poor results for our regression tasks. We append \textbf{-linear} to each method to indicate usage of linear regression. To obtain even stronger baselines, we apply deep learning to both feature vectors and graph kernels, indicated by \textbf{-deep}. % A deep neural network is expected to learn higher-level representations from the raw input.

\textbf{Smoothed graph kernels}. Yanardag et al.~\cite{yanardag2015structural} apply smoothing to graph kernels, which extends their method of deep graph kernels~\cite{yanardag2015deep} by considering structural similarity between sub-structures. We report smoothed results only on deep neural networks as it outperforms  alternatives empirically.

\textbf{PSCN}, which applies convolutional neural networks (CNN) to locally connected regions from graphs~\cite{niepert2016learning}, achieving better results over graph kernels on some of the classification data sets.

\textbf{Hyper-parameters}. All hyper-parameters are tuned to obtain the best results on validation set. For linear regression, we chose the L2-coefficient from $\{10^{0}, 10^{-1}, ..., 10^{-7}\}$. For neural network regression, the initial learning rate is selected from $\{0.1,$ $0.05,$ $0.01,$ $...,$ $10^{-4}\}$, the number of hidden layers from $\{1, 2, ..., 4\}$, and the hidden layer size from $\{32, 64, ..., 1024\}$. The size of the graphlets for GK is chosen from $\{3, 4\}$ (higher than 4 is extremely slow), the height parameter of WL from $\{2, 3, 4\}$, the discount parameter for smoothed graph kernels from $\{1, 0.8, ..., 0\}$. Following~\cite{niepert2016learning} for PSCN, the width is set to the average number of nodes, and the receptive field size is chosen between 5 and 10.

\textbf{Notes.} Please notice that in our experiments we are not identifying the nodes in the networks or using the information of the nodes outside the network itself. Of course, knowing the president of United States is in the network provides more confidence on its growth. We choose not to identify nodes because (1) this study focuses on investigating the predictive power of the topological structure of networks, and (2) in practice information about individual nodes may not be available for privacy reasons. For the same reasons, we do not include any information other than the network structure (e.g., content of tweets, or historical metrics of the network) in the prediction task, even though including more information may improve the prediction accuracy. 

It is also worth mentioning that even though most networks in our data sets are subgraphs of a much larger network (e.g., the Facebook friendship network), we only use the structure within the subgraphs and do not touch the outside structure of the global network. This is because partitioning a large network into subgraphs is just a way to create abundant networks for training and testing the model. In reality we may make predictions of an entire network, or the global structure outside a network may not be observable.

\subsection{DeepGraph Model Parameters}
\paragraph*{\bf Hyper-parameters and Preprocessing} 
\begin{comment}
Specifically, each column of $K$ is first centralized by the mean and variance of this column, computed over all training examples. Based on the centralized distribution, histograms can be calculated. By denoting $N_B$ the number of bins used in the histogram, we obtain a descriptor as a matrix $S \in \mathds{R}^{N_B \times N}$. Unlike adjacent matrix representation, the HKS histogram representation is independent of vertex ordering and the number of vertexes.  
\end{comment}
For parameters included in HKS, we set them to default values across all data sets without further tuning. In Equation~\ref{equ:HKS}, we set $t_1 = 0.1$, $t_N=25$, and $N=64$. Number of bins $N_B$ is set to 64. 
%We choose 64 for both $N$ and $N_B$ to avoid margin issues for filters of CNN. 
In order to compute histograms, HKS values above $+1.2$ and below $-1.2$ standard deviation are respectively put to the first and last bins. Values in between are assigned to the remaining equally divided 62 bins.

We perform standard normalization for the histograms of graphs. Each histogram is preprocessed by pixel-wise normalization. We compute the mean and standard deviation for each pixel over the training data set. Then each pixel is normalized by subtracting the corresponding mean value and being divided by sd\footnote{$\epsilon=10^{-8}$ is added to the denominator to avoid numeric issues.} .

We initialize the parameters of the neural networks using a Gaussian distribution with zero mean and unit standard deviation. An adaptive optimizer, Adam, is used to optimize the parameters of the neural networks. 
%, which is an adaptive stochastic gradient optimization method to train deep neural networks. 
Default hyper-parameters of Adam are used~\cite{kingma2015adam}.%: a discount factor of 0.9 to compute the accumulated discount sum of the first moment vector of gradients and 0.999 for the second order moment vector and 0.001 as the learning rate. Note that we do not apply domain/dataset-specific hyper-parameter tuning. The same training procedure is applied to all the data sets. 

Structure related hyper-parameters of DeepGraph is set to be the same across datasets. There are two multiresolution convolution layers for each network column, with number of filters 32 and 16. For each convolution layer, we apply three sizes of filters, which are 2, 4, and 6. TanH is used as the activation function. There are two fully connected layers both of size 256. Dropout is applied to the last two dense layers with probability of 0.5. Other learning parameters are listed in Table~\ref{tab:hyperparam_setup}.

\begin{table}[h!]
\caption{Setup of hyper-parameters for DeepGraph.}
\begin{center}
\small
\setlength{\tabcolsep}{2pt}
\begin{tabular}{c|p{1.1cm}|c|c|c|c} 
\hline 
 & Facebook & YouTube & AAN & IMDB & Weibo \\
\hline 
L2-coefficient & \multicolumn{1}{|c|}{1e-5} & 1e-5 & 0.005 & 1e-5 & 1e-6 \\
\hline 
Init learning rate & \multicolumn{1}{|c|}{0.005} & 0.01 & 5e-4 & 0.005 & 0.001 \\
\hline 
\end{tabular}
\end{center}
\label{tab:hyperparam_setup}
\end{table}

\subsection{Variants of DeepGraph}
To assign the credit of each key component in our DeepGraph model, 
%In addition to apply multicolumn mulitresolution convolution network to our graph descriptor, 
we also experiment with some of its variants, by feeding our graph descriptor (\textbf{GD}) to a linear regressor (\textbf{GD-linear}), a standard convolutional neural network (\textbf{GD-CNN}), and a multilayer perceptron (\textbf{GD-MLP}). Hyper-parameters for these models are tuned similarly as baselines.
\section{Experiment results}
\label{sec:exp}

%In this section, we present the results of the experiments as designed in Section 4. 

\subsection{Overall performance}
\begin{table}[h!]
\caption{Performance measured by MSE (the lower the better), where original label $y$ is scaled to $\log_2(y+1)$.}
\begin{center}
\small
\setlength{\tabcolsep}{2pt}
\begin{tabular}{c|c|c|c|c|c} 
\hline 
Dataset & Facebook & YouTube & AAN & IMDB & Weibo \\
\hline 
Feature-deep & 1.107  & 2.623  & 0.421  & 0.527  & 2.244  \\
Feature-linear & 1.116  & 2.633  & 0.439  & 0.525  & $2.346^{***}$ \\
\hline 
GK-smooth & $1.313^{***}$ & $2.675^{***}$ & $0.480^{***}$ & $0.561^{**}$ & $3.011^{***}$ \\
GK-deep & $1.315^{***}$ & $2.671^{***}$  & $0.492^{***}$  & $0.565^{**}$ & $3.061^{**}$ \\
GK-linear & $1.335^{***}$ & $2.736^{***}$  & $0.519^{**}$ & $0.576^{***}$ & $3.242^{***}$ \\
\hline 
WL-smooth & $1.158^{***}$ & $2.659$ & $0.434$ & $0.536$ & $2.397^{***}$ \\
WL-deep & $1.165^{**}$ & $2.654$ & $0.437$ & $0.532$ & $2.403^{***}$ \\
WL-linear & $1.331^{***}$ & $2.702^{***}$ & $0.445$ & $0.596^{***}$ & $2.411^{***}$ \\
\hline 
SP-smooth & 1.138 & 2.615  & 0.422  & 0.530  & $2.315^{***}$ \\
SP-deep & $1.155^{**}$ & 2.607  & 0.428  & $0.531$ & $2.322^{***}$  \\
SP-linear & $1.179^{***}$ & $2.613$ & 0.432  & $0.535$ & $2.359^{***}$ \\
\hline 
PSCN & 1.117  & $2.534^{***}$ & 0.425  & 0.528  & $2.441^{***}$ \\
\hhline{=|=|=|=|=|=}
\multicolumn{1}{l|}{\tiny Proposed methods}&&&&&\\
GD-linear & $1.174^{***}$ & $2.750^{***}$ & $0.587^{***}$ & $0.583^{***}$ & $2.812^{***}$ \\
GD-MLP & $1.082^{*}$ & $2.427^{***}$ & $0.394^{***}$ & $0.513^{*}$ & $2.080^{***}$ \\
GD-CNN & 1.087  & $2.429^{***}$ & $0.391^{***}$ & $0.512^{*}$ & $2.114^{***}$ \\
DeepGraph & \boldmath$1.068^{**}_{\triangledown}$ & \boldmath$2.409^{***}_{\triangledown\triangledown}$ & \boldmath$0.379^{***}$ & \boldmath$0.508^{***}_{\triangledown}$ & \boldmath$1.961^{***}_{\triangledown\triangledown\triangledown}$ \\
\hline 
\end{tabular}

\flushleft ``***(**)" means the result is significantly better or worse over \textit{Features-dp} according to paired t-test test at level 0.01(0.1). ``$\scriptstyle\triangledown$'' means DeepGraph-multi is better than the better one between DeepGraph-MLP and DeepGraph-CNN.
\end{center}
\label{tab:mse_results}
\end{table}

The overall performance of all competing methods across data sets are displayed in~\ref{tab:mse_results}. We make the following observations. First, integrating graph descriptor with deep learning, our method DeepGraph outperforms all competing methods significantly. This empirically confirms that graph descriptor could preserve more information of the network structure than bag-of-substructures, both globally and locally. In contrast, utilizing manually designed features could lead to loss of information.

GD-MLP and GD-CNN have already gain improvement over the strongest baseline on most of the data sets, while DeepGraph can further improve the performance by utilizing the semantics of HKS-based graph descriptor. 
%the multicolumn mulitresolution convolution network. 
This shows that we can indeed extract more useful features by applying column-wise and row-wise convolution over graph descriptors.

Comparing with GD-linear, which applies linear regression on top of the HKS-based graph descriptor, DeepGraph, GD-MLP, and GD-CNN performs significantly better. This indicates that the effectiveness of the HKS-based graph descriptor has to be utlized by a ``deeper'' model which explores the convolutions and non-linear transformations of the low-level representation. %The plain GraphDescriptor outperforms other network representation methods only on some of the data sets, when simple linear regression is applied. This results in the simplicity of 
%A simple linear learner fails to generalize abstract information from the raw output of graph descriptors. This proves the successfulness of applying deep learning techniques, which abstracts high-level, latent representations from raw out put of graph descriptors. %The two comparisons together verify the power of both graph descriptors and deep learning. 

%It is interesting to notice that GraphDescriptor alone perform poorly when trained with linear regression. This indicates that the effectiveness of the HKS representation has to be utlized by a ``deeper'' model which explores the convolutions and non-linear transformations of the low-level representation. %The plain GraphDescriptor outperforms other network representation methods only on some of the data sets, when simple linear regression is applied. This results in the simplicity of 
%A simple linear learner fails to generalize abstract information from the raw output of graph descriptors. %This is clearer when we look at Figure~\ref{fig:hks_examples}: Figure~\ref{fig:hks_examples} (a) and (b) have similar descriptors, but perturbation exists. It is hard for a linear regressor to recognize the similarity between (a) and (b) but this is not a problem for a convolution network.

%Second, bag-of-substrcutures can be improved largely when trained through deep learning techniques. Deep-Quads easily outperforms all methods trained by linear regression.

Comparing feature based methods with other baselines, the former exhibit strong prediction power. Incorporating both local and global information of the networks, the hand-crafted features are very indicative of the network growth, which is hard for automatic methods to compete.

When trained on deep networks, the performance of graph kernels could be improved over their linear version. Smoothing kernels can further bring in some improvement. By applying convolution over locally connected regions of the graphs, PSCN can beat many graph kernels on most data sets. These results are consistent with previous studies~\cite{niepert2016learning,yanardag2015structural}.

\subsection{Computational Cost of DeepGraph}
Training of DeepGraph is very fast. The models are converged in less than 10 minutes on a Titan X GPU.  % to converge. Here we mainly focus on the computation of graph descriptors. 
The major overhead of DeepGraph is the computation of the HKS-based graph descriptors. 
We empirically measure the computation time for all data sets on a server with 2.40 GHz CPU and 120G RAM. The graphs in our data sets have size as large as 5,000 nodes and 200,000 edges, which is enough for most network prediction problems~\cite{kupavskii2012prediction,romero2011interplay,yang2010modeling}. The generation of graph descriptors takes an average of 0.86 hour per data set. In contrast, the strongest baseline, feature based method, takes 7.9 hours on average to generate all features. While the strongest graph kernel, SP, takes nearly 5 days.

\subsection{Feature Analysis}
\begin{figure}[t]
\begin{minipage}[t]{0.44\textwidth}
\centering
\begin{minipage}[t]{0.44\textwidth}
\centering
\includegraphics[width=0.90\textwidth]{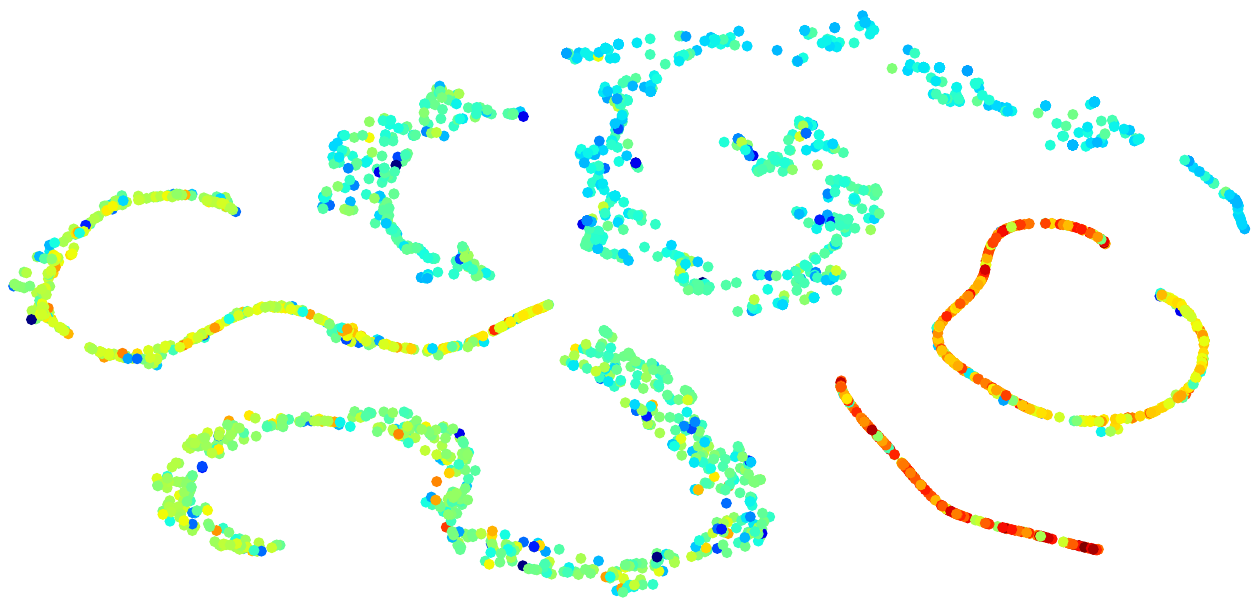}\\
(a) \# edges.
\end{minipage}
\begin{minipage}[t]{0.44\textwidth}
\centering
\includegraphics[width=0.98\textwidth]{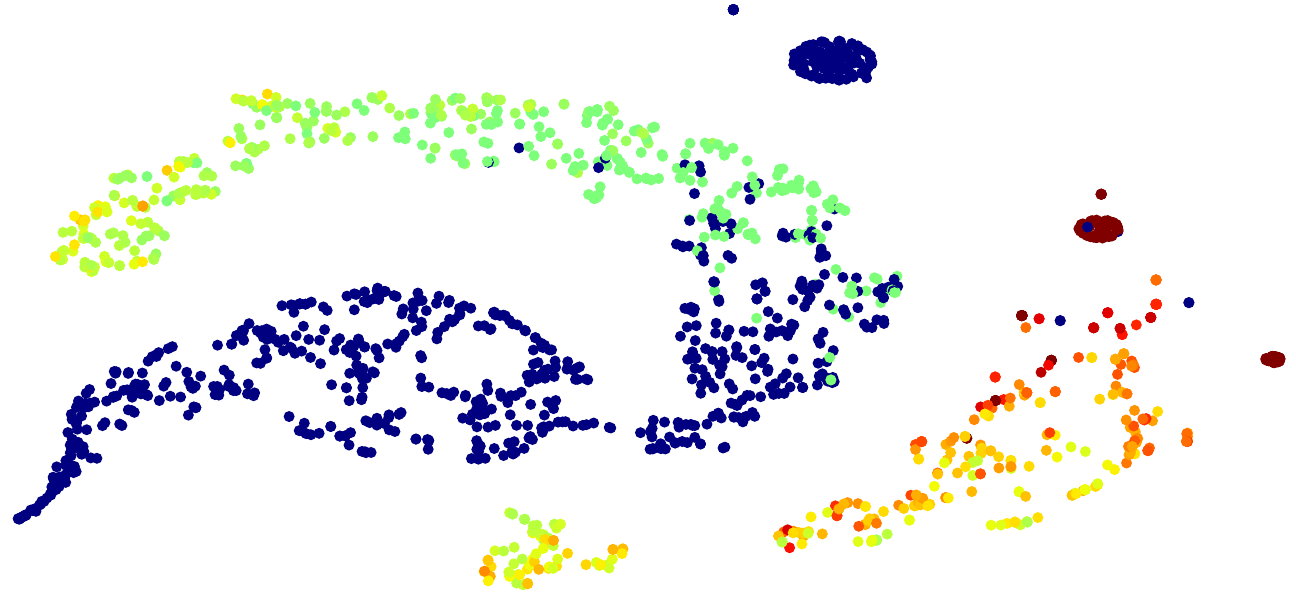}\\
(b) \# closed triangles.
\end{minipage}
\begin{minipage}[t]{0.06\textwidth}
\centering
\includegraphics[width=0.9\textwidth]{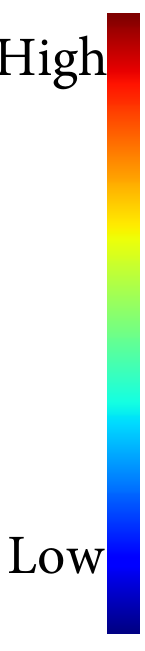}\\
\end{minipage}
\end{minipage}
\begin{minipage}[t]{0.44\textwidth}
\centering
\begin{minipage}[t]{0.44\textwidth}
\centering
\includegraphics[width=0.90\textwidth]{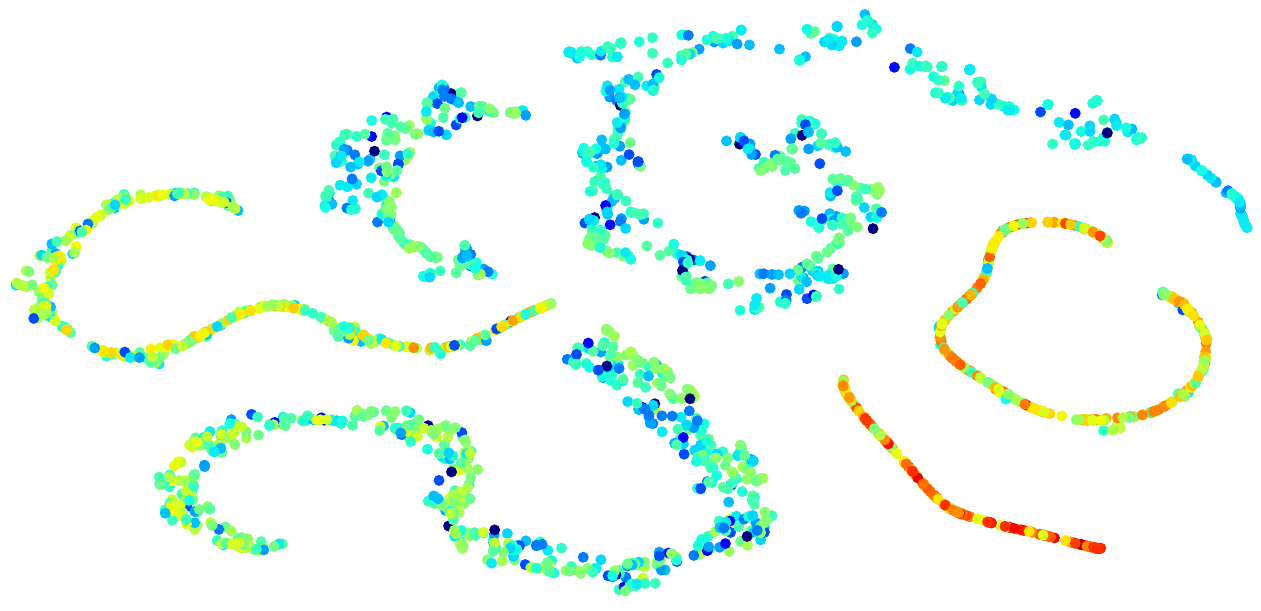}\\
(c) \# leaf nodes.
\end{minipage}
\begin{minipage}[t]{0.44\textwidth}
\centering
\includegraphics[width=0.98\textwidth]{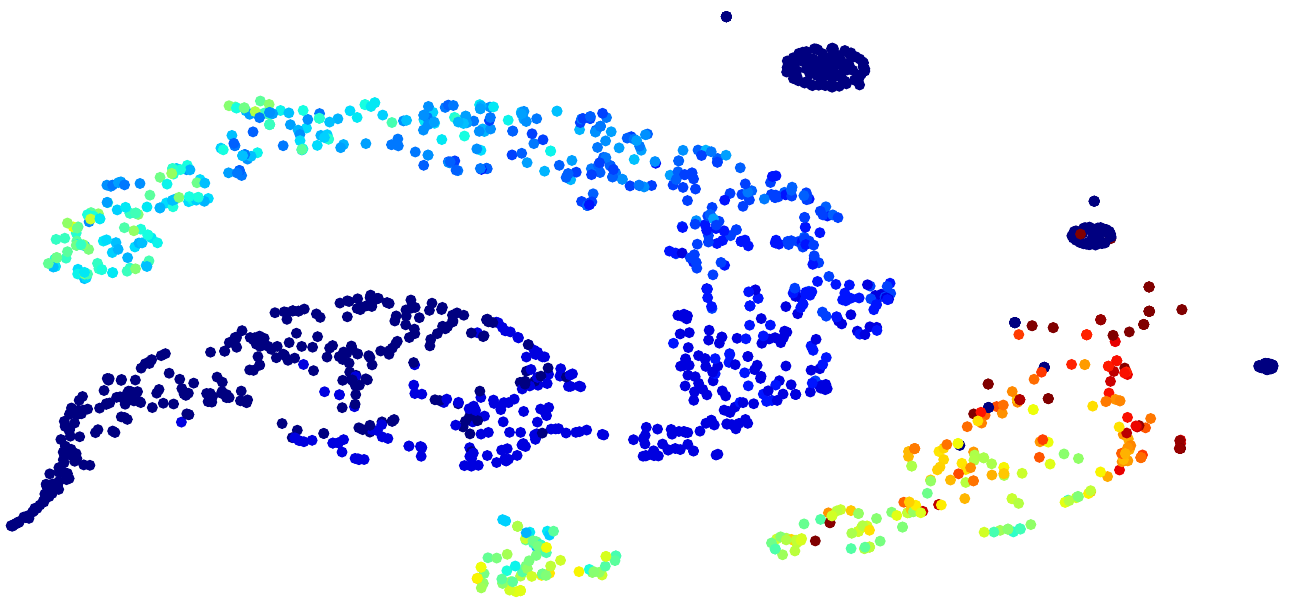}\\
(d) \# open triangles.
\end{minipage}
\end{minipage}
\begin{minipage}[t]{0.44\textwidth}
\centering
\begin{minipage}[t]{0.44\textwidth}
\centering
\includegraphics[width=0.90\textwidth]{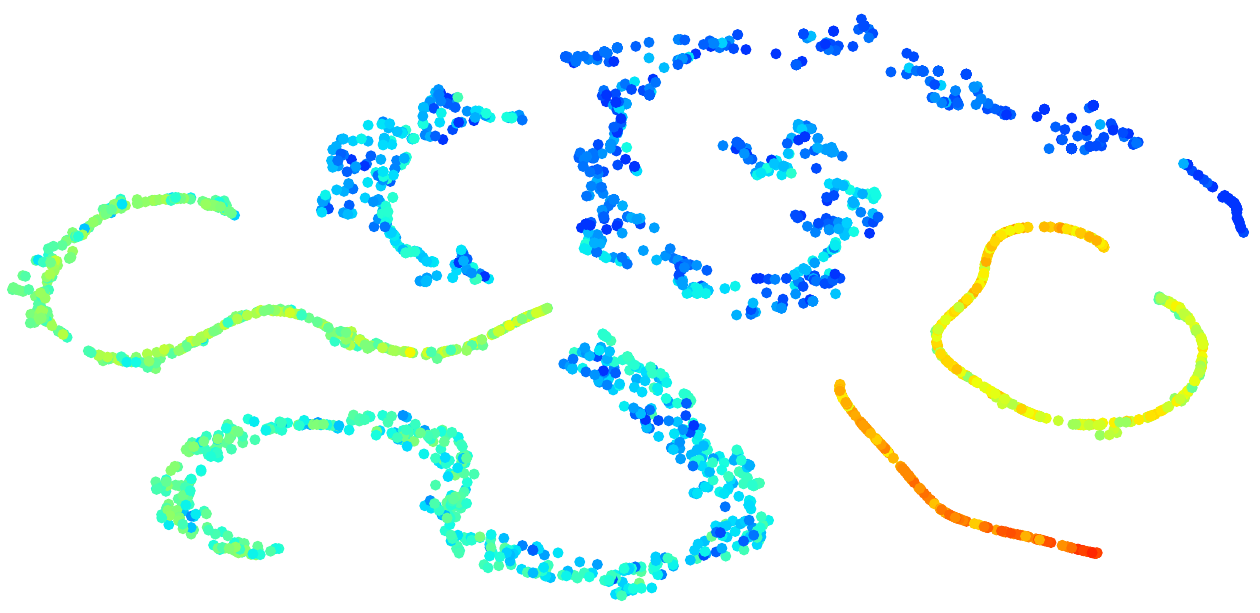}\\
(e) \# communities.
\end{minipage}
\begin{minipage}[t]{0.45\textwidth}
\centering
\includegraphics[width=0.98\textwidth]{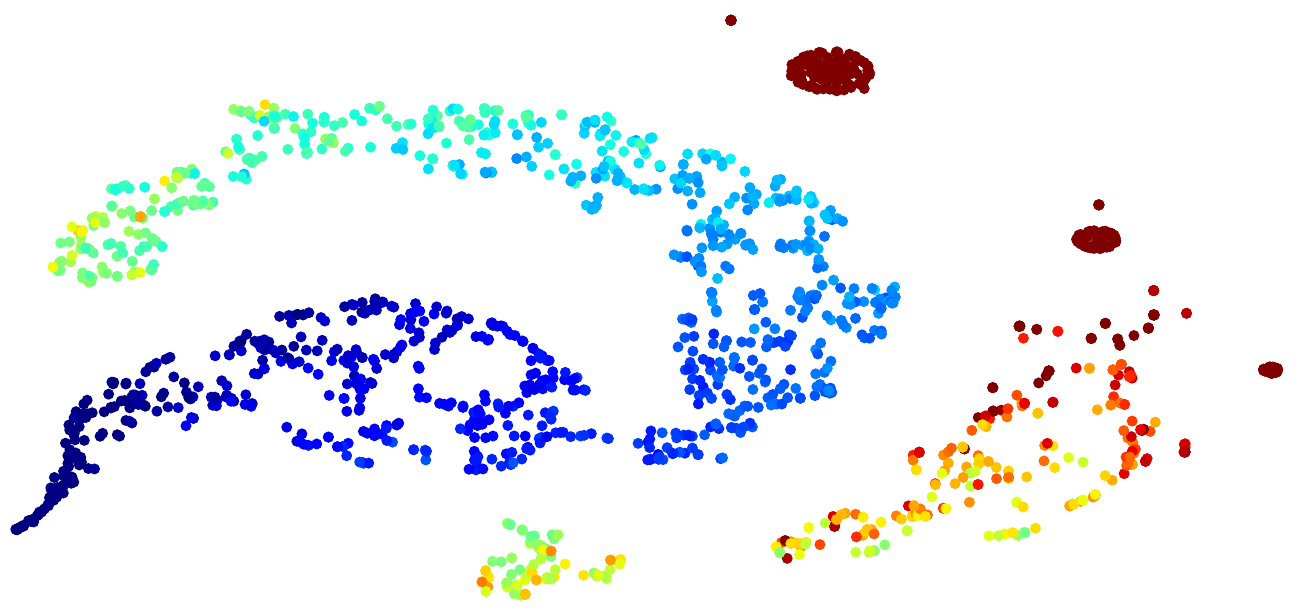}\\
(f) Edge density.
\end{minipage}
\end{minipage}
\begin{minipage}[t]{0.44\textwidth}
\centering
\begin{minipage}[t]{0.44\textwidth}
\centering
\includegraphics[width=0.90\textwidth]{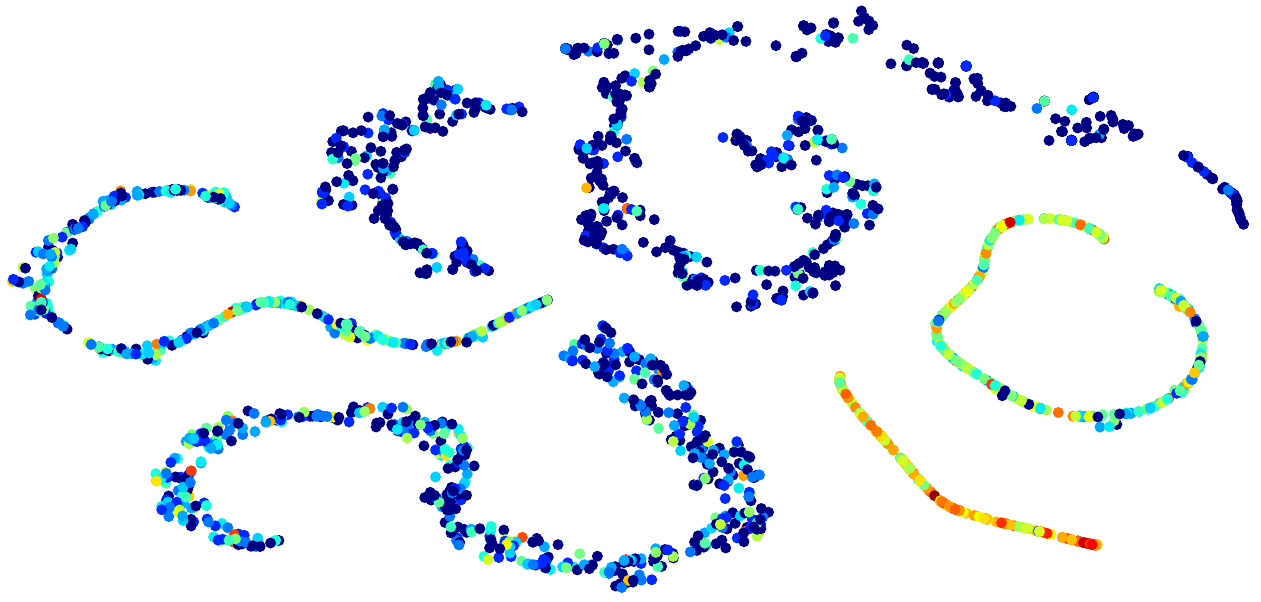}\\
(g) Growth (Size).
\end{minipage}
\begin{minipage}[t]{0.44\textwidth}
\centering
\includegraphics[width=0.98\textwidth]{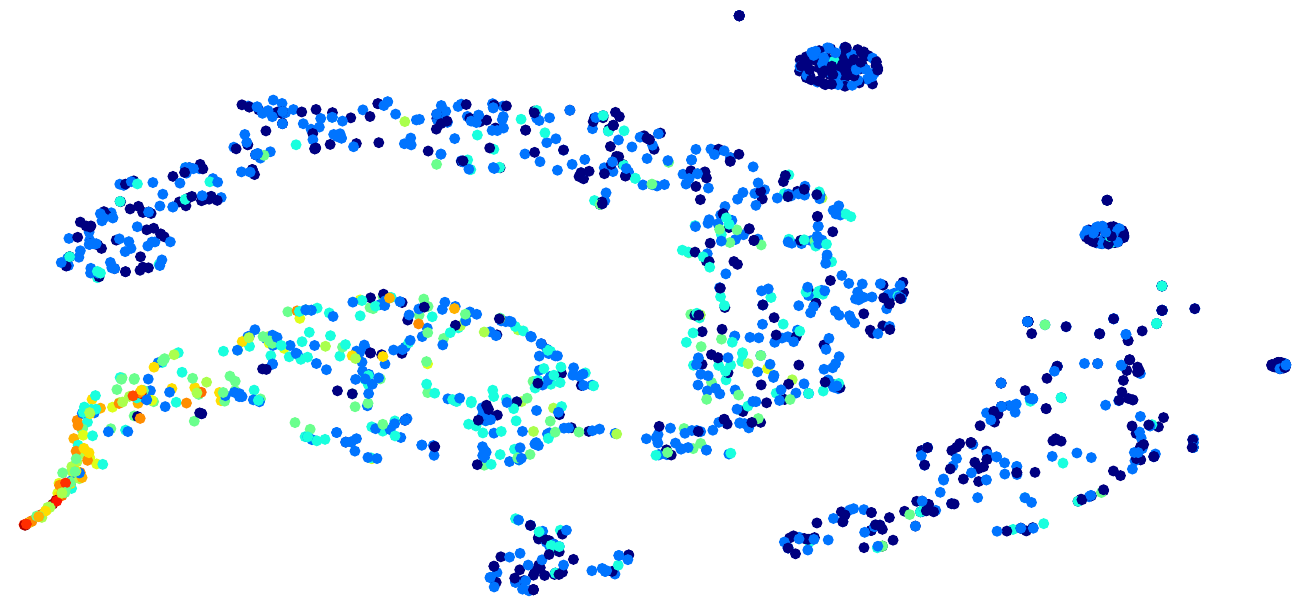}\\
(h) Growth (h-index).
\end{minipage}
\caption{Feature visualization. One point is a graph in \textit{test} set. The layout is produced from high-level representations of DeepGraph, colored using structural, \textit{hand-crafted} network properties, which are presented under each subfigures. Red (blue) color indicates high (low) property values. The left column displays graphs from Weibo, while the right column shows AAN. The bottom row displays the ground-truth growth of the network properties.}
\label{fig:feature_analysis}
\end{minipage}
\end{figure}

It has been shown empirically that DeepGraph could well abstract high-level features to represent graphs. It is intriguing to know whether these learned features correspond to well-known structural patterns in network literature. To this end, we select some of the network properties manually computed for the feature based method. Note that we work only on test set, as we care more about the prediction performance. These properties characterize either global or local aspects of networks, and are listed in Figure~\ref{fig:feature_analysis}. 
%including the number of edges, network density, the number of open and closed triangles, the number of leaf nodes, and the number of communities.

%In order to examine whether the high-level representations learned by DeepGraph have captured these properties, we need a way to visualize the high-level representations and the above network properties. To do so, 
The feature vectors output by the last hidden layer of DeepGraph are fed to t-SNE~\cite{blondel2008fast}, a dimensionality reduction algorithm for visualizing high-dimensional data sets. 
%The t-SNE algorithm projects feature vectors into a 2-dimensional space, where similar vectors are projected closely. 
The visualizations of data sets Weibo and AAN are displayed in Figure~\ref{fig:feature_analysis}. 
%For graphs of Weibo, all graphs lie continuously, forming a smooth manifold. For AAN, three clusters are formed. 
We obtain similar results on other data sets, which are omitted to conserve space.

To connect the hand-crafted structural properties with the learned high-level features, we color individual graphs by the values of these properties (e.g., network density). Patterns on the distribution of colors could suggest a connection between learned features and the network property.

As we could observe, properties on left column of Figure~\ref{fig:feature_analysis} exhibit some related patterns. For example, in the Figure 5(e), the graphs clustered to the top have the fewest communities, while graphs in the bottom right corner have the most. This is interesting as an ego-net with a larger number of communities implies that the graph center lies in between bigger communities, which are likely to be structural holes in the global network. According to social network literature \cite{burt2000network}, nodes spanning structural holes are likely to gain social capital, promoting the growth of its ego-net. Indeed, when we compare the color scheme of 5(e) with 5(g), which plots the actual growth of the network sizes, we can see that the number of communities in a diffusion network is indeed positively correlated with the growth of the diffusion network.  

Top three figures on the right column also show some common characteristics: the left cluster forms a band, with lower values on the bottom and higher ones on the top. Graphs with higher values mostly cluster to the right area.

These common patterns suggest that the high-level features output by deep learning have indeed captured these network properties. As we include both local and global properties, we demonstrate from another perspective that DeepGraph could learn global-to-local structural information from the network topology. Comparing to 5(h), which plots the actual growth of the h-index of the ego-nodes, we can see the correlation of the features are much weaker than those in the left column (Weibo network). This is not surprising, as h-index is a property that can not be directly derived from the network itself, and thus the prediction task is much harder. 

Some additional observations can be made from Figure~\ref{fig:feature_analysis}. First, as the number of open and closed triangles are actually features of graphlets~\cite{shervashidze2009efficient,ugander2013subgraph}, we can see that DeepGraph has automatically learned these useful features without human input. Second, since edge density is a function of the number of edges and nodes, DeepGraph not only learns the number of edges and nodes (we do not show the node property in Figure~\ref{fig:feature_analysis}, but this is true), but also their none-linear relationship that involves division.
%Moreover, this relationship is not simply a linear combination of the two numbers. Rather, non-linear transformations like division and squaring are involved.

% Theoretically, the major overhead of computing graph descriptors lies in the calculation of eigenvectors. Its time complexity can be reduced to $O(K|V|)$ by computing only the first $K$ eigenvectors and eigenvalues.

\subsection{Error Analysis}
%Analyzing prediction errors could bring in more insight into the behaviors of methods. Ideally, we would like to look into the graphs where our methods make errors, and summarize commonalities from these graphs. However, 
Graphs in our data sets typically have hundreds of nodes, which is hard for humans to directly generalize useful information from a set of graphs. As a compromise, we characterize graphs by a set of simple network properties, e.g., the number of nodes, edges, and edge density.

We first want to investigate graphs for which DeepGraph makes more mistakes than baseline, and also the other way around. Here we use the strongest baseline, feature-based method as our reference. The procedure is as follows: among graphs where DeepGraph has smaller MSE than the baseline, we select the top 100 with the largest MSE differences between the two methods. For these top graphs, we compute the average of the properties mentioned above. Similar procedure is also applied to the baseline.

%\begin{table}[h!]
%\caption{Statistics of graphs for which DeepGraph outperforms Deep-Quad (first rows), and vise versa (second rows). The third row of each data set shows the average statistics on its test set.}
%\small
%\begin{center}
%\setlength{\tabcolsep}{3pt}
%\begin{tabular}{|c|c|c|c|c|c|} 
%\hline 
%Data set & Method & \# nodes & \# edges & Avg. degree & Edge density \\
%\hline 
%Facebook & DeepGraph & 742.60 & 17002.00 & 21.24 & 0.034 \\
% & Deep-Quad & 621.35 & 10324.50 & 14.81 & 0.030 \\
% & Avg. & 401.74 & 6461.31 & 11.74 & 0.052 \\
%\hline 
%YouTube & DeepGraph & 1014.10 & 4410.30 & 6.84 & 0.017 \\
% & Deep-Quad & 447.80 & 1784.40 & 6.08 & 0.027 \\
% & Avg. & 157.93 & 1042.93 & 3.82 & 0.068 \\
%\hline 
%AAN & DeepGraph & 994.35 & 7135.60 & 13.75 & 0.037 \\
% & Deep-Quad & 602.65 & 4223.65 & 13.03 & 0.052 \\
% & Avg. & 467.39 & 6569.37 & 12.12 & 0.097 \\
%\hline 
%IMDB & DeepGraph & 208.45 & 3936.60 & 33.72 & 0.219 \\
%& Deep-Quad & 185.05 & 2842.30 & 27.59 & 0.230 \\
% & Avg. & 160.33 & 6314.32 & 37.38 & 0.379 \\
%  \hline 
% Weibo & DeepGraph & 312.75 & 396.60 & 2.19 & 0.031 \\
% & Deep-Quad & 232.70 & 184.00 & 1.63 & 0.015 \\
%  & Avg. & 125.19 & 137.80 & 1.70 & 0.053 \\
%\hline 
%\end{tabular}
%\end{center}
%\label{tab:error_analysis_compare}
%\end{table}

The statistics of graphs where either DeepGraph or the baseline significantly outperforms the other are higher than the average statistics of each data set. This could result form the skewed distribution of the data set -- a large number of graphs are of smaller size, leading to more training instances of small graphs. 
%Moreover, both methods perform reasonably well on smaller graphs so the variance of performance is small. 
We also observe that both methods perform reasonably well on denser networks. % and there is a larger variance of performance on sparse networks. 

On the other hand, graphs on which DeepGraph performed better have relatively larger sizes than those where the baseline performed better. This indicates that the HKS representation has an advantage on larger graphs, the structures of which are more difficult to be represented by a bag of local substructures.

\section{Conclusion}
\label{sec:conclusions}
We present a novel neural network model that predicts the growth of network properties based on its graph structure. This model, DeepGraph, computes a new representation of the graph structure based on heat kernel signatures. A multi-column, multi-resolution convolution neural network is designed to further learn the high-level representations and predict the network growth in an end-to-end fashion. Experiments on large collections of real-world networks prove that DeepGraph significantly outperforms methods based on hand-crafted features, graph kernels, and competing deep learning methods. The higher-level representations learned by DeepGraph well correlate with findings and theories in social network literature, showing that a deep learning model can automatically discover meaningful and predictive structural patterns in networks.

Our study reassures the predictive power of network structures and suggests a way to effectively utilize this power. A meaningful future direction is to integrate network structure with other types of information, such as the content of information cascades in the network. A joint representation of multi-modal information may maximize the performance of particular prediction tasks. 

%
% The following two commands are all you need in the
% initial runs of your .tex file to
% produce the bibliography for the citations in your paper.
\bibliographystyle{abbrv}
\newpage
\bibliography{survey}  % sigproc.bib is the name of the Bibliography in this case
% You must have a proper ".bib" file
%  and remember to run:
% latex bibtex latex latex
% to resolve all references
%
% ACM needs 'a single self-contained file'!
%
%\balancecolumns
\end{document}